\documentclass[aps,prb,showpacs,reprint]{revtex4-1}

\usepackage{graphicx}
\usepackage{amsmath}
\usepackage{bbm}

\begin{document}
\title
{Effect of current hysteresis on the spin polarization of current in
a paramagnetic resonant tunneling diode}
\author{P. W\'ojcik}
\author{B.J. Spisak}
\email[Electronic address: ]{spisak@novell.ftj.agh.edu.pl}
\author{M. Wo{\l}oszyn}
\author{J. Adamowski}
\affiliation{Faculty of Physics and Applied Computer Science, AGH
University of Science and Technology, al. Mickiewicza 30, 30-059
Krak\'ow, Poland}

\begin {abstract}
A spin-dependent quantum transport is investigated in a paramagnetic
resonant tunneling diode (RTD) based on a Zn$_{1-x}$Mn$_x$Se/ZnBeSe
heterostructure. 
Using the Wigner-Poisson method and assuming the two-current model we
have calculated the current-voltage characteristics, potential energy
profiles and electron density distributions for spin-up and spin-down
electron current in an external magnetic field. 
We have found that -- for both the spin-polarized currents -- two types
of the current hysteresis appear on the current-voltage characteristics.
The current hysteresis of the first type occurs at the bias voltage
below the resonant current peak and results from the accumulation of
electrons in the quantum well layer. 
The current hysteresis of the second type appears at the bias voltage
above the resonant current peak and is caused by the creation of the
quasi-bound state in the left contact region and the resonant
tunneling through this quasi-bound state.  
The physical interpretation of both the types of the current hysteresis
is further supported by the analysis of the calculated self-consistent
potential profiles and electron density distributions. 
Based on these results we have shown that -- in certain bias voltage and
magnetic field ranges -- the spin polarization of the current exhibits
the plateau behavior with the nearly full spin polarization. 
This property is very promising for possible applications in
spintronics.
\end{abstract}


\maketitle

\section{Introduction}\label{sec:intro}
A rapid progress in homo- and heteroepitaxy of semiconductors with
magnetic 
dopants~\cite{Ohno1992,Ohno1998,Hayashi2001,Edmonds2002,Yu2002,Chiba2003,Ku2003,Eid2005,Jungwirth2005}
has led to the interest in the transport properties of the diluted
magnetic semiconductors (DMSs) due to their potential spintronic
applications.  
A special attention is paid to the studies of
(Zn,Mn)Se~\cite{Furdyna1988,Yu1995} and
(Ga,Mn)As.\cite{Ando2003,Liu2005,Jungwirth2006,Dietl2001}.
The nanostructures consisting of the non-magnetic and magnetic 
layers~\cite{Gruber2001,Slobodskyy2003,Beletskii2003} lead to a
possibility of creation of the nanodevice, in which spin polarization 
of the current can be controlled by the bias voltage.
The fabrication of an effective spin filter for application to
semiconductor spintronic nanodevices, such as spin
transistor~\cite{Datta1990}, is one of the most important challenge
for the contemporary nanotechnology.

The application of DMS nanostructure as an effective spin filter has
been first proposed by Egues.~\cite{Egues1998} 
The nanostructure studied in Ref.\cite{Egues1998} consisted of a
paramagnetic semiconductor layer  made of Zn$_{1-x}$Mn$_x$Se, sandwiched
between two non-magnetic ZnSe layers. 
In the presence of the external magnetic field, the giant Zeeman effect
occurring in the Zn$_{1-x}$Mn$_x$Se layer~\cite{Furdyna1988} causes that
the paramagnetic layer acts as a potential well for spin-down  electrons
and as a potential barrier for spin-up electrons, which leads to the
total current dominated by the spin-down electrons.
In this type of the nanodevice, the spin polarization of the current
can be controlled by the external magnetic field.
There is another type of the nanodevice, namely, the resonant tunneling 
diode (RTD) with the paramagnetic quantum well, in which the spin
polarization of the current can be controlled by the bias voltage in
the presence of the external magnetic field.
If the magnetic field is applied to the paramagnetic RTD, the resonant
tunneling conditions are satisfied for the different bias voltages for
the spin-up and spin-down electrons due to the giant Zeeman splitting
of the quasi-bound state in the paramagnetic quantum well.
This leads to the separation of the corresponding resonant current
peaks and consequently to the spin polarization of the current.
The paramagnetic RTD based on the ZnBeSe/ZnMnSe heterostructure
was experimentally demonstrated by Slobodskyy~et~al.,~\cite{Slobodskyy2003}
and theoretically investigated by Havu~et~al.~\cite{Havu2005}
In both the experimental and theoretical studies, the separation of the 
resonant current peaks corresponding to the different spin components of
the current was found. 
Nevertheless, in these studies, we did not find any hint for a current
hysteresis and its influence on the spin polarization of the current.
On the other hand, the nonlinear effects, in particular, the current 
hysteresis (current bistability), can be observed in the resonant
tunneling through the nanodevices made from the non-magnetic
materials.\cite{Goldman1987,Sheard1988,Sollner1987}
The first observation of the current hysteresis in the nonmagnetic 
(AlGa)As/GaAs/(AlGa)As heterostructure was reported by Goldman and
Tsui.\cite{Goldman1987}
The current hysteresis\cite{Goldman1987} occurs in the negative 
differential resistance (NDR) region of the current-voltage
characteristics and is attributed to the accumulation of the electrons 
in the quantum well layer.\cite{Sheard1988}  
Sollner~\cite{Sollner1987} proposed another explanation of the current
bistability based on the oscillations of a circuit containing an element
that exhibits the NDR.
In non-magnetic RTDs, the following two types of the current hysteresis 
can be distinguished: (I) the hysteresis occurring in the bias voltage 
range below the resonant current peak~\cite{Su1991,Macks1996}, (II) the
hysteresis occurring in the bias voltage range above the resonant 
current peak.\cite{Goldman1987b}
Moreover, one can observe both the types of hysteresis on the same 
current-voltage characteristics (double hysteresis).\cite{Goldman1987}
Recently, Dai et al.~\cite{Dai2006} have investigated a possibility of 
tuning the current hysteresis in the non-magnetic resonant tunneling
diode by changing the geometric parameters of the nanodevice and found
that the increase of the collector barrier width can enhance the current
hysteresis of type (I), while the decrease of the collector barrier 
width enhances the type (II) current hysteresis.
These results~\cite{Dai2006} indicate that the geometric parameters 
considerably affect the nonlinear transport properties of the
nanodevice.

However, in the paramagnetic RTD, the current hysteresis and its 
influence on the spin polarization of the current has not been studied
until now.
In this paper we present the results of such investigations.
We show that the current hysteresis is of crucial importance for 
obtaining the spin-polarized current.
We provide the physical interpretation of this effect based on the
analysis of the self-consistent potential profile and the electron 
density distribution.
We have predicted the bistable behavior of the spin current
polarization as a function of the bias voltage and the appearance of
plateaus of the spin current polarization as a function of the
external magnetic field.

The paper is organized as follows: in Sec. 2, we describe the model of 
paramagnetic RTD and the self-consistent Wigner-Poisson method. 
Section 3 contains the results, Section 4 -- the discussion, and Section
5 -- the conclusions and summary.

\section{Model and Theory}\label{sec:model}

\begin{figure}
\includegraphics[width=\columnwidth]{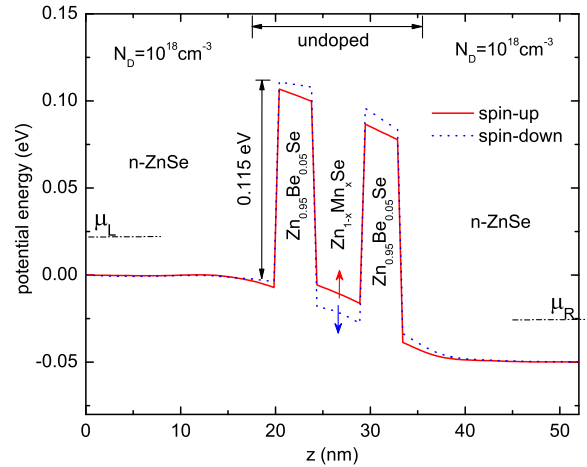}
\caption{\label{fig:1}(Color online) Potential energy profile in the
paramagnetic RTD for the electrons with spin-up (solid line, red) and 
spin-down (dotted line, blue).
Coordinate $z$ is measured along the layer growth direction, 
$\mu_{L(R)}$ is the electrochemical potential of the left (right)
contact. 
The undoped (active) region consists of the paramagnetic quantum well
made of Zn$_{1-x}$Mn$_{x}$Se sandwiched between two
Zn$_{0.95}$Be$_{0.05}$Se potential barrier layers.
The active region of the nanodevice is separated from the $n$-doped 
ZnSe contacts by the two spacer layers at the left and right side.}
\end{figure}

We consider the paramagnetic RTD diode consisting of the 
Zn$_{1-x}$Mn$_x$Se paramagnetic layer sandwiched between the two
Zn$_{0.95}$Be$_{0.05}$Se layers (Fig.~\ref{fig:1}). 
The active region of the nanodevice is separated from $n$-doped ZnSe
contacts by two spacer layers located at the left and right contacts.
The difference between the conduction band minima of Zn$_{1-x}$Mn$_x$Se 
and Zn$_{0.95}$Be$_{0.05}$Se leads to the potential energy profile with
two barriers and one quantum well, in which the quasi-bound states can 
be formed.
We focus on the effect of bias voltage $V_b$ applied between the left 
and right contacts and the external magnetic field $\mathbf{B} = (0,0,B)$ 
applied in the growth direction on the spin polarization of the current.
In the presence of the magnetic field, the exchange interaction between
the spins of the Mn$^{2+}$~ions and the spins of the conduction band
electrons leads to the giant Zeeman splitting of the conduction band
minima in the paramagnetic quantum well.\cite{Furdyna1988} 
This splitting occurs at temperatures below the Curie temperature of
Zn$_{1-x}$Mn$_x$Se and gives rise to the different potential energy
profiles for spin-up and spin-down electrons.
For the small concentration of Mn$^{2+}$ ions the giant Zeeman 
splitting can be expressed by the formula\cite{Furdyna1988}
\begin{equation}
\Delta (B)= \frac{1}{2}N_0\alpha \, x\,S_0 \,B_S
\left( \frac{g \mu _BSB}{k_B T_{\mathrm{eff}}}
\right ) \; ,
\label{eq:GZS}
\end{equation}
where $N_0\alpha=0.26$ eV is the sp-d exchange constant for the 
conduction band electrons,\cite{Twardowski2005} $x$ is the concentration
of Mn$^{2+}$ ions, $S_0B_S$ is the effective Brillouin function for spin
$S=5/2$ that corresponds to the spin of Mn$^{2+}$ ion, $g$ is the
effective Land{\'e} factor, $\mu _B$ is the Bohr magneton, $S_0$ and
$T_{\mathrm{eff}}$ are the phenomenological parameters accounting for
Mn-Mn interaction that are chosen as follows:\cite{Yu1995}  $S_0=1.60$
and $T_{\mathrm{eff}}=1.5$~K for $x=0.043$, while $S_0=1.18$ and
$T_{\mathrm{eff}}=2.55$~K for $x=0.083$.

The potential energy profiles for the spin-up and spin-down electrons 
in the nanodevice are displayed in Fig.~\ref{fig:1}.

We calculate the current-voltage characteristics of the paramagnetic 
RTD using the modified version of the self-consistent
Wigner-Poisson approach.
For this purpose we extend our previous approach\cite{Wojcik2010} by
assuming the two-current model,\cite{bookZiman2000,Spisak2009} according
to which the conduction band electrons are described by the
spin-dependent Wigner distribution function (WDF).
In order to calculate the electronic transport through the nanodevice 
we apply the simplified version of the quantum kinetic
equation.\cite{Weng2003,Spisak2005,Spisak2009}
Assuming the translational invariance in the $x-y$ plane, the quantum 
transport equations for the steady state can be reduced to the following
one-dimensional form:
\begin{equation}
 \frac{\hbar k}{m}\frac{\partial \rho^w_{\sigma}(z,k)}{\partial z}
 = \frac{i}{2\pi \hbar}\int{\mathrm{d}k^{\prime}}~U^w_{\sigma}
(z,k-k^{\prime}) \rho^w_{\sigma}(z, k^{\prime}) \;,
\label{eq:WEq}
\end{equation}
where $\rho^w_{\sigma}(z,k)$ is the spin-dependent WDF, $k$ is the
$z$-component of the wave vector, $m$ is the electron conduction band
mass, and $\sigma=\uparrow, \downarrow$ is the electron spin index.

The non-local Wigner potential for spin channel $\sigma$  is given by 
the formula
\begin{eqnarray}
 U^w_{\sigma}(z,k-k^{\prime}) & = &
\int{\mathrm{d}z^{\prime}}~\big[U_{\sigma}(z+z^{\prime}/2)-
\label{eq:npot}\\*
 & - & U_{\sigma}(z-z^{\prime}/2)\big]
 \exp{\big[-i(k-k^{\prime})z^{\prime}\big]} \;, \nonumber
\end{eqnarray}
where $U_{\sigma}(z)$ is the total spin-dependent potential energy that 
consists of the two terms: conduction band potential energy
$U_{\sigma}^0(z,B)$ and Hartree potential energy $U_{\sigma}^H(z)$.

At this stage of our study, we neglect the exchange interaction between 
the conduction band electrons.
The conduction band potential energy has the form
\begin{eqnarray}
U_{\sigma}^0(z;B) & = & \sum_{i=1}^N U_i \Theta(z-z_i)\Theta(z_{i+1}-z) 
+ \label{eq:cpot}\\*
& + & \Delta(B) \Theta(z-z_1)\Theta(z_2-z) \;, \nonumber
\end{eqnarray}
where $z_i$ is the position of the barrier/well 
(paramagnetic/non-magnetic) interface, $\Theta(z)$ is the Heaviside
step function, $U_i$ is the height of the $i$-th barrier and
$\Delta(B)$ is the giant Zeeman splitting of the conduction band in the
paramagnetic layer.

In Eq.~(\ref{eq:cpot}), the Zeeman splitting for the conduction band 
electrons has been neglected because it is a few orders of magnitude
smaller than the giant Zeeman splitting occurring in the paramagnetic
quantum well.

The Hartree potential energy satisfies the Poisson equation
\begin{equation}
\frac{\mathrm{d}^2U_{\sigma}^H(z)}{\mathrm{d}z^2}=
\frac{e^2}{\varepsilon_0\varepsilon}[N_D(z)-n_{\sigma}(z)] \;,
\label{eq:PEq}
\end{equation}
where $\varepsilon_0$ is the vacuum electric permittivity, $\varepsilon$
is the relative static electric permittivity, and $N_D(z)$ is the
concentration of the ionized donors.

The density of the electrons with spin $\sigma$ can be expressed as 
follows:
\begin{equation}
n_{\sigma}(z)=\frac{1}{2\pi}\int{\mathrm{d}k}\, \rho_{\sigma}^w(z,k) \;.
\label{eq: ed}
\end{equation}
Quantum kinetic equation (\ref{eq:WEq}) and Poisson equation 
(\ref{eq:PEq}) constitute a system of equations that can be solved by a
self-consistent numerical procedure.\cite{Kim2007}
We assume the Dirichlet boundary conditions for the Poisson equation,
i.e., $U_{\sigma}^H(0)=0$ and $U_{\sigma}^H(l)=-eV_b$, where $V_b$ is
the bias voltage applied between the left (L) and right (R) electrodes
separated by distance $l$.

When solving quantum kinetic equation~(\ref{eq:WEq}) we use the open 
boundary conditions\cite{Frensley1990} in the form
\begin{eqnarray}
\rho_{\sigma}^w(0,k)\bigg|_{k>0}&=&f_{L\sigma}(k) \; , \\ \nonumber
\rho_{\sigma}^w(l,k)\bigg|_{k<0}&=&f_{R\sigma}(k) \; ,
\label{eq:bc}
\end{eqnarray}
where $f_{\nu\sigma}(k)\; (\nu=L,R)$ is the supply 
function\cite{bookFerry2009} that for contact $\nu$ has the form
\begin{equation}
f_{\nu\sigma}(k)=\frac{mk_BT}{\pi \hbar^2}
\ln{\bigg\{1+\exp{\bigg[-\frac{1}{k_BT}
\bigg(\frac{\hbar^2k^2}{2m}-\mu_{\nu\sigma}\bigg)}\bigg]\bigg\}} \;,
\label{eq:sf}
\end{equation}
where $T$ is the temperature, and $\mu_{\nu\sigma}$ is the
electrochemical potential of reservoir $\nu$.

The current density for non-interacting spin channels can be expressed 
by the following  formula:
\begin{equation}
j_{\sigma}(V_b,B)=\frac{e}{2\pi}\int{\mathrm{d}k}\,\frac{\hbar k}{m}
\rho_{\sigma}^w(z,k;V_b,B) \;.
\label{eq:fc}
\end{equation}
We note that although the right-hand side of Eq.~(\ref{eq:fc}) contains 
coordinate $z$, the current is independent of  $z$ for the steady state
solution.  
The present calculations have been performed with the geometric 
parameters of the nanodevice, for which the most pronounced current
hysteresis has been found in non-magnetic RTD.\cite{Dai2006}
We have checked that -- in the paramagnetic RTD -- the current 
hysteresis effects are also the strongest for the same geometry as in
nonmagnetic RTD.  
In the nanodevice studied in the present paper, the contacts are made
from the $n$-type ZnSe with the homogeneous ionized donor concentration
$N_D=2\times10^{18}$ cm$^{-3}$, the thickness of each contact is equal 
to 17~nm, the thickness of each spacer layer is 3~nm, the thickness of
each potential barrier (well) layer is 3~nm (5~nm), the total length of
the nanodevice $l = 54$~nm, and the height of the potential barrier
$U_0 = 0.115$~eV.\cite{Chauvet2000} 
The energy of the left-contact conduction band bottom is taken as the
reference energy and set equal to 0.   
Due to the small thickness of the double-barrier region we assume that
the electrons are described by the conduction band effective mass of
ZnSe, i.e., $m=0.16\,m_0$, where $m_0$ is the free electron mass.
We take on the relative electric permittivity $\varepsilon=8.6$ and the 
lattice constant $a=0.5667$~nm for ZnSe.
The present simulations have been carried out for temperature $T=1.2$ K.
We have applied the computational grid with $N_z=95$ mesh points for 
coordinate $z$ and $N_k=72$ mesh points for wave vector $k$.

\section{Results}\label{sec:results}

\begin{figure*}
 \includegraphics[width=0.5\textwidth]{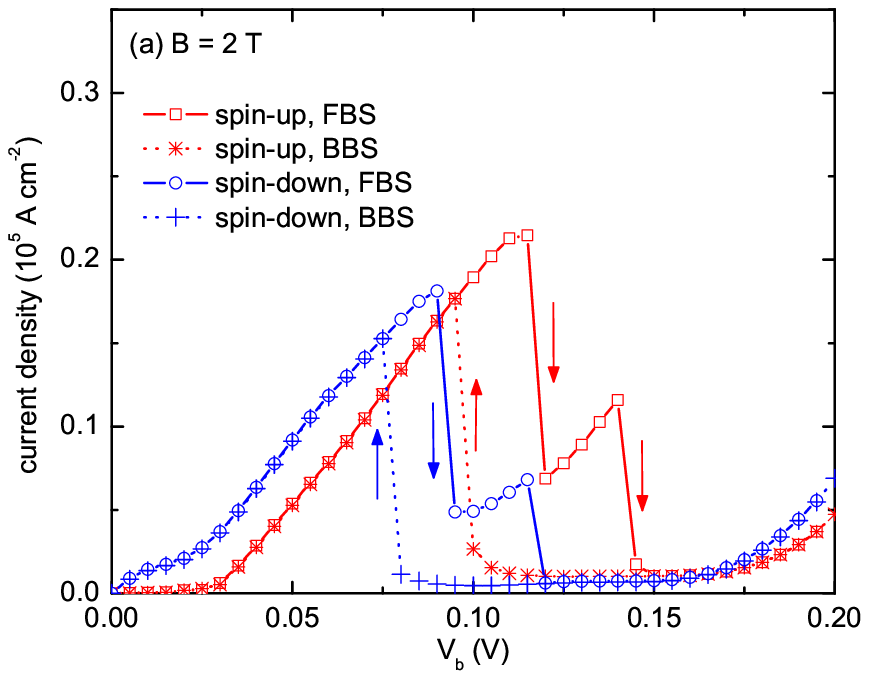}%
 \includegraphics[width=0.5\textwidth]{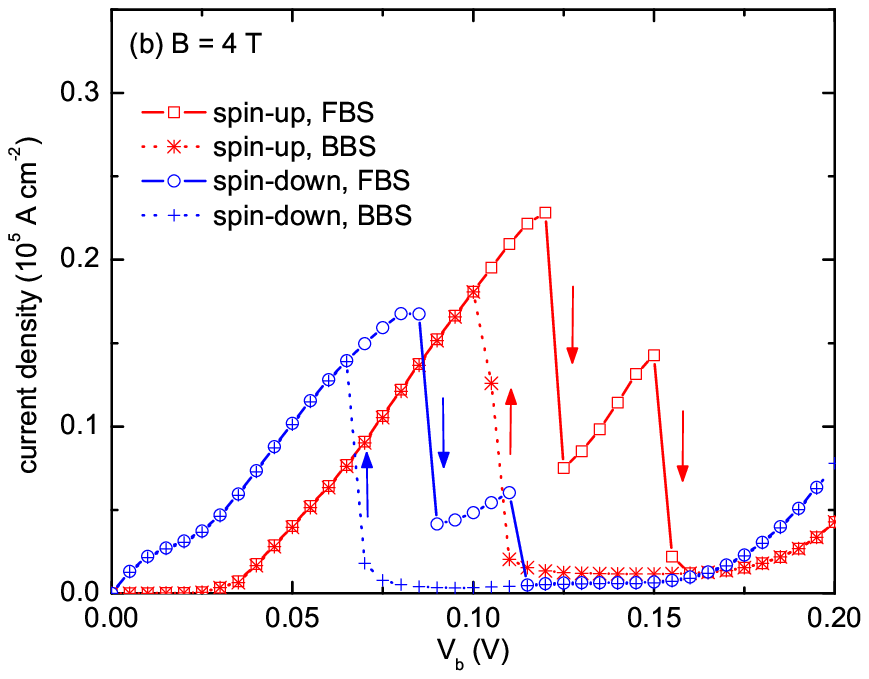}\\
 \includegraphics[width=0.5\textwidth]{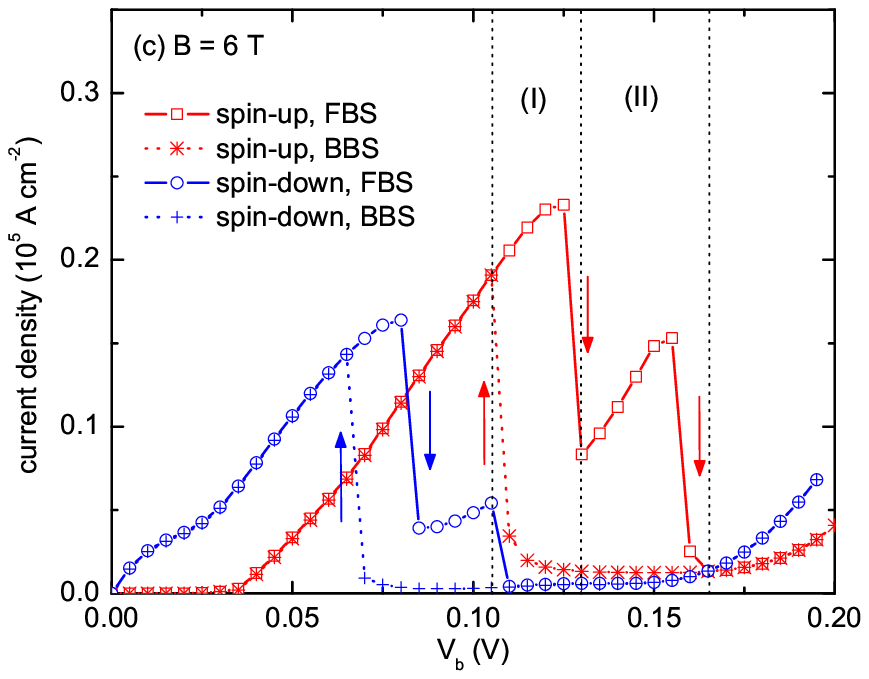}%
 \includegraphics[width=0.5\textwidth]{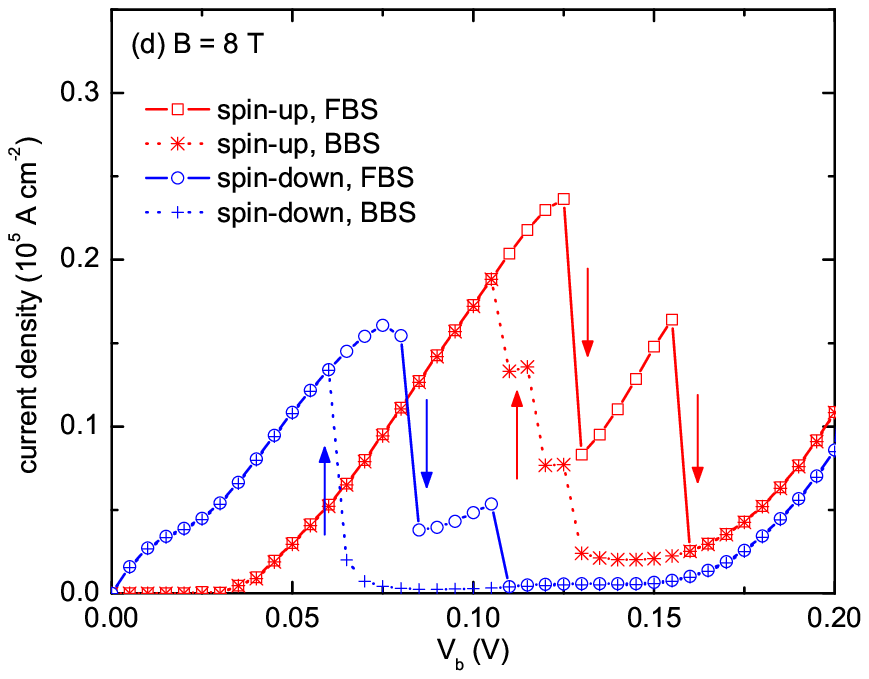}
 \caption{\label{fig:2}(Color online) Current-voltage characteristics 
of the paramagnetic RTD for spin-up (red)  and spin-down (blue) currents
for magnetic field (a) $B=2$~T, (b) $B=4$~T, (c) $B=6$~T, and (d)
$B=8$~T. 
The results for the forward (FBS) and backward (BBS) bias sweep are
displayed by solid line and dotted curves, respectively. 
The directions of the bias sweep are marked by the arrows. 
In panel (c), the first (I) and the second (II) bias voltage range, in 
which the current hysteresis occurs, are marked.}
\end{figure*}

Using the steady state Wigner-Poisson approach, described in 
Sec.~\ref{sec:model}, we have performed calculations of current-voltage
characteristics $I(V_b)$ for the paramagnetic RTD in the presence of
external magnetic field for the spin-up and spin-down current 
components, separately. 
Each of these characteristics have been obtained for the two directions
of the bias sweep: forward bias sweep (FBS), in which the bias voltage
increased from $0$~V to $0.2$~V with step $0.005$~V and backward bias
sweep (BBS), in which the bias voltage decreased from $0.2$~V to $0$~V
with step $0.005$~V.
In Fig.~\ref{fig:2}, we present the calculated current-voltage 
characteristics for different values of the magnetic fields.

\begin{figure}
 \includegraphics[width=\columnwidth]{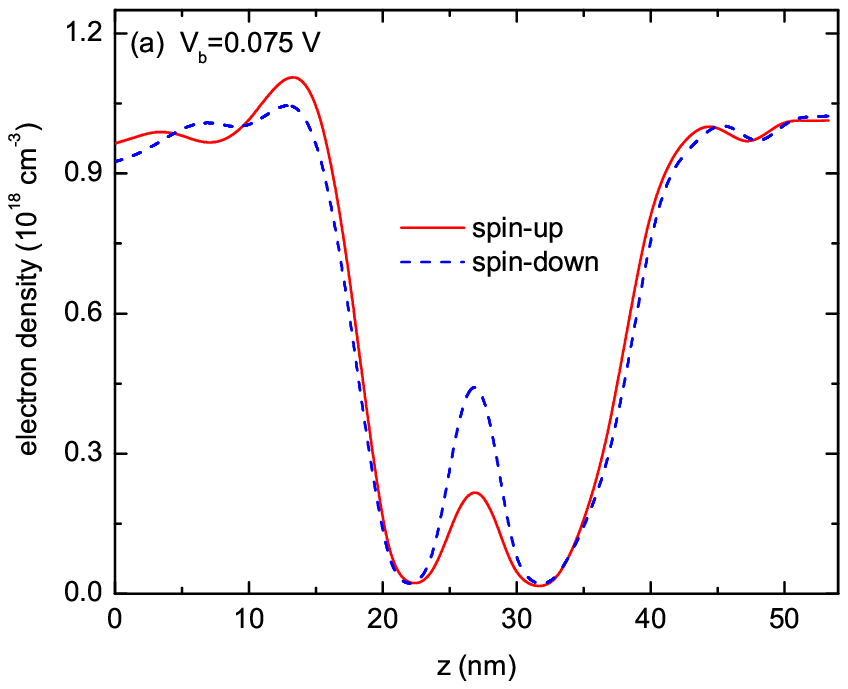}\\
 \includegraphics[width=\columnwidth]{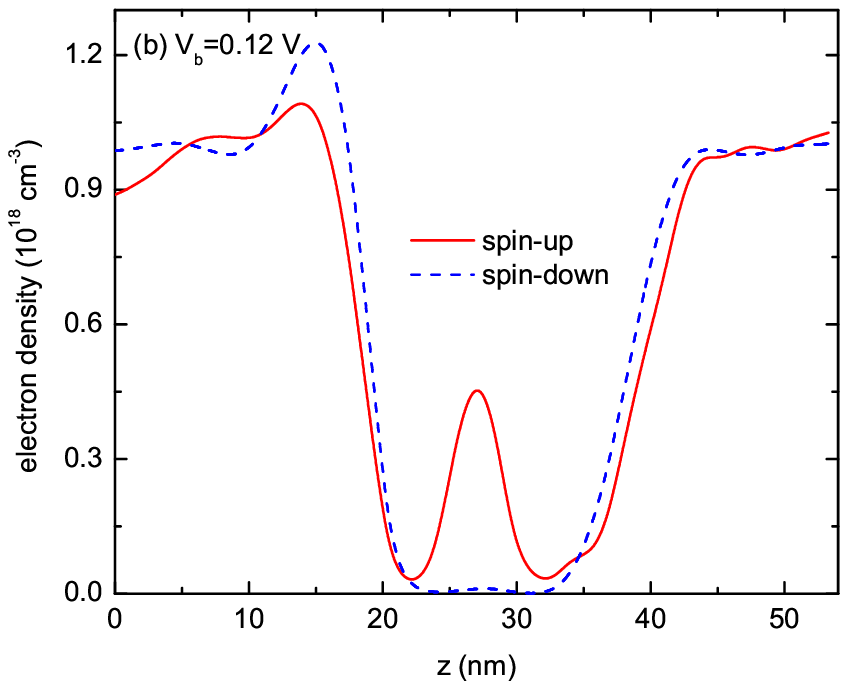}
 \caption{\label{fig:3}(Color online) Electron density distribution for 
FBS at magnetic field $B=6$~T.
 Panel (a) shows the results for $V_b=0.075$~V that corresponds to the 
resonant current peak for spin-down electrons,  panel (b) shows the
results for $V_b=0.12$~V that corresponds to the resonant current peak
for spin-up electrons.}
\end{figure}

The results of Fig. \ref{fig:2} show that the resonant current peaks 
corresponding to the spin-up and spin-down current components for FBS
and BBS are separated. 
The separation of the resonant current peaks results from the giant
Zeeman splitting of the energy levels corresponding to the quasi-bound
states formed in the paramagnetic quantum well and increases with the
increasing magnetic field.
This means that the resonance conditions for the spin-up and spin-down 
electron currents are satisfied for the different values of the bias
voltage. 
In order to get a more deep physical insight into this effect, we
present in Fig.~\ref{fig:3} the spin-dependent electron density for
magnetic field $B=6$~T in the case of FBS and for the following two
values of the bias voltage: $V_b=0.075$~V [Fig.~\ref{fig:3}~(a)] that
corresponds to the current peak for spin-down current component and
$V_b=0.12$~V [Fig.~\ref{fig:3}~(b)] that corresponds to the current peak
for spin-up current component.
The larger accumulation of the spin-down electrons in the quantum well 
[Fig.~3(a)] indicates that the resonance condition is better satisfied
for these electrons.  
In Fig.~3(b), the electron density distributions for spin-up and
spin-down electrons are completely different in the quantum well layer.
In this case, we have obtained the fairly high concentration of spin-up 
electrons with a simultaneous almost complete depletion of spin-down
electrons in the quantum well.
This electron density distribution corresponds to the resonance 
conditions  satisfied for the spin-up electrons only.

\begin{figure}
\includegraphics[width=\columnwidth]{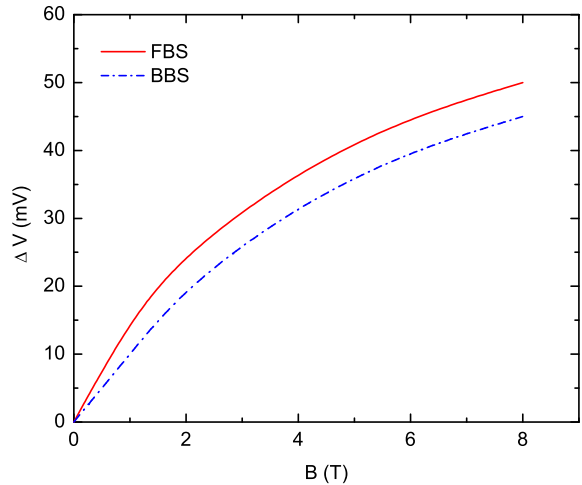}
\caption{\label{fig:4}(Color online) Separation $\Delta V$ between the 
resonant current peaks for the spin-up and spin-down current components
as a function of magnetic field $B$ in the case of FBS (solid, red) and
BBS (dashed, blue).}
\end{figure}

The results plotted in Figs.~\ref{fig:2}(a-d) show that the increasing 
magnetic field shifts the position of the current maximum for the
spin-up current component towards the  higher values of the bias
voltage, while the position of the current maximum for the spin-down
current component is shifted towards the lower values of the bias
voltage.
Fig.~\ref{fig:4} displays the separation 
$\Delta V=V^{peak}_{up}-V^{peak}_{down}$ between the resonant current
peaks for the spin-up and spin-down current components as a function of
magnetic field.
The separation between these peaks increases with the increasing 
magnetic field and reaches $\sim 50$~mV at $B=8$~T.
For each value of the magnetic field $\Delta V$ for BBS is lower than 
that for FBS.

The efficiency of the spin filter is characterized by the spin 
polarization of the current defined as follows:
\begin{equation}
P_j=\frac{j_{\uparrow}-j_{\downarrow}}{j_{\uparrow}+j_{\downarrow}} \:,
\label{eq:P}
\end{equation}
where $j_{\uparrow}$ and $j_{\downarrow}$ are the current densities for 
the spin-up and spin-down electrons, respectively.

\begin{figure}
 \includegraphics[width=\columnwidth]{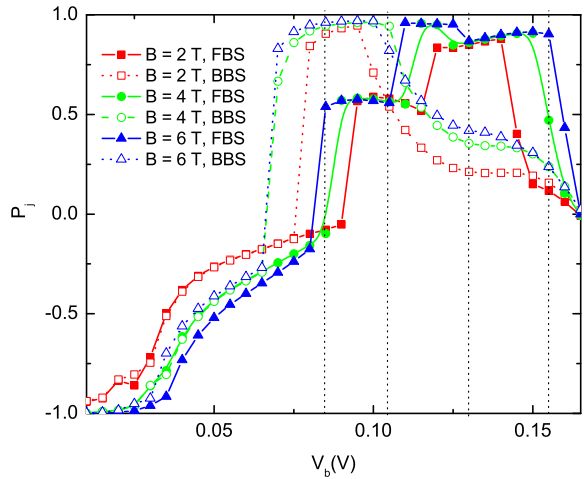}
 \caption{\label{fig:5}(Color online) Spin polarization $P_j$ of the 
current as a function of bias voltage $V_b$ for the FBS and BBS  and for
different values of magnetic field $B$.  
The bias voltage ranges corresponding to the plateaus of the spin
current polarization are marked by the vertical dotted lines.}
\end{figure}

\begin{figure}
 \includegraphics[width=\columnwidth]{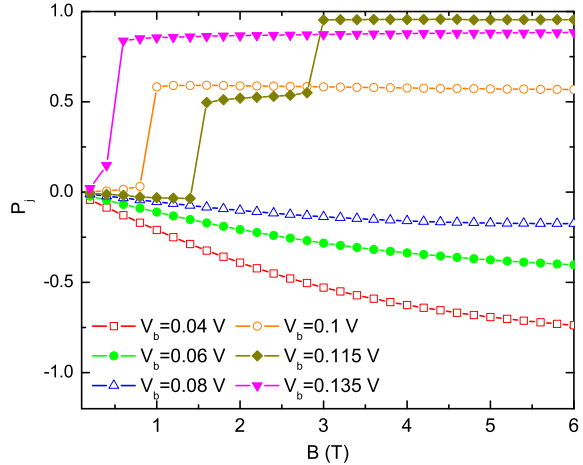}
 \caption{\label{fig:6}(Color online) Spin polarization $P_j$ of the 
current as a function of magnetic field $B$ for the FBS  and different
bias voltage $V_b$.}
\end{figure}

Fig.~\ref{fig:2} shows that the current-voltage characteristics for 
the spin-up and spin-down current components obtained for the increasing
bias (FBS) essentially differ from those obtained for the decreasing
bias (BBS).
This leads to the hysteresis that occurs for each spin component of the
current.
These hysteresis loops and the phenomena responsible for them determine
the current polarization.
The calculated spin current polarization is depicted in Fig.~\ref{fig:5}
as a function of bias voltage, and Fig.~\ref{fig:6} as a function of
magnetic field.  
We observe the distinct plateau regions of the spin current polarization
that appear for certain bias voltage (Fig.~\ref{fig:5}) and magnetic
field (Fig.~\ref{fig:6}).
In some ranges of the bias voltage and magnetic field, we have obtained
almost full spin polarization of the current flowing through the
paramagnetic RTD.
The different spin current polarization obtained for FBS and BBS 
[Fig. \ref{fig:5}] results from the current hysteresis (cf.
Fig.~\ref{fig:2}).
For example, for $V_b=0.1$~V and $B=6$~T the spin current polarization 
$P _j\simeq 1$ in the case of the BBS but $P_j \simeq 0.5$ in the case
of the FBS.
Fig.~\ref{fig:5} shows that the spin current polarization can be
large, i.e., $0.8\leq P_j \leq 1$, and  the spin-up current component
dominates for both the FBS and BBS.
In the case of the BBS, the bias voltage range corresponding to the
almost full spin current polarization is shifted toward the lower
voltage.
The closer inspection of the $P_j(V_b)$ curve for the FBS shows that we 
have obtained the plateau behavior of the spin current polarization in
quite wide bias voltage ranges [cf. Fig.~\ref{fig:5}].
The width of these ranges increases with the increasing magnetic field.
In Fig.~\ref{fig:5}, the bias voltage ranges corresponding to the
plateaus of the spin current polarization are marked by the vertical
dotted lines.

The plateau behavior of the spin current polarization also occurs
if we change the magnetic field keeping the bias voltage fixed.
This effect is demonstrated on Fig.~\ref{fig:6}, which displays the
results of calculations for the FBS.
Depending on the bias voltage we obtain the dominating current of the 
spin-down electrons for 0 V $\leq V_b \leq$ 0.08 V, while for $V_b >$
0.08 V the spin-up electrons give the main contribution to the current.
While the spin-down current polarization monotonically increases with
the increasing magnetic field, the spin-up current polarization rapidly
increases at low magnetic field and becomes almost constant in the range
of the intermediate and high magnetic field.
In Fig.~\ref{fig:6}, we can distinguish the following ranges of the
magnetic field, in which the plateau behavior of the spin current
polarization occurs: from 1~T to 6~T for $V_b=0.1$~V and from 0.6~T to
6~T for $V_b=0.135$~V.
For $V_b=0.115$~V we have obtained the three magnetic-field ranges with
almost constant spin current polarization: the first from 0~T to 1.6~T
with $P_j \simeq 0$, the second from 1.6~T to 3~T with $P_j \simeq 0.5$,
and the third from 3~T to 6~T with $P_j \simeq 0.95$.

\section{Discussion}\label{sec:discusion}

The calculated current-voltage characteristics [Fig.~\ref{fig:2}]
show the hysteresis for each spin component of the current.
We have found that the current hysteresis occurs in the following two 
separate bias voltage ranges: (I) the first one with the bias voltage
values below that the corresponding to the resonant current peak and
(II) the second with the bias voltage values above that corresponding to
the resonant current peak [cf. Fig.~\ref{fig:2}(c)].
In these two ranges, the current hysteresis originates from two 
different effects.
In region (I), the current hysteresis  results from the accumulation of
the electron charge in the central quantum well, i.e., that made from
the Zn$_{1-x}$Mn$_x$Se [cf. Fig.~\ref{fig:1}].
In region (II), the current hysteresis originates from the creation of
the quasi-bound state in the quantum well induced in the region of the
left contact and the resonant electron tunneling through this state.

\begin{figure}
\includegraphics[width=\columnwidth]{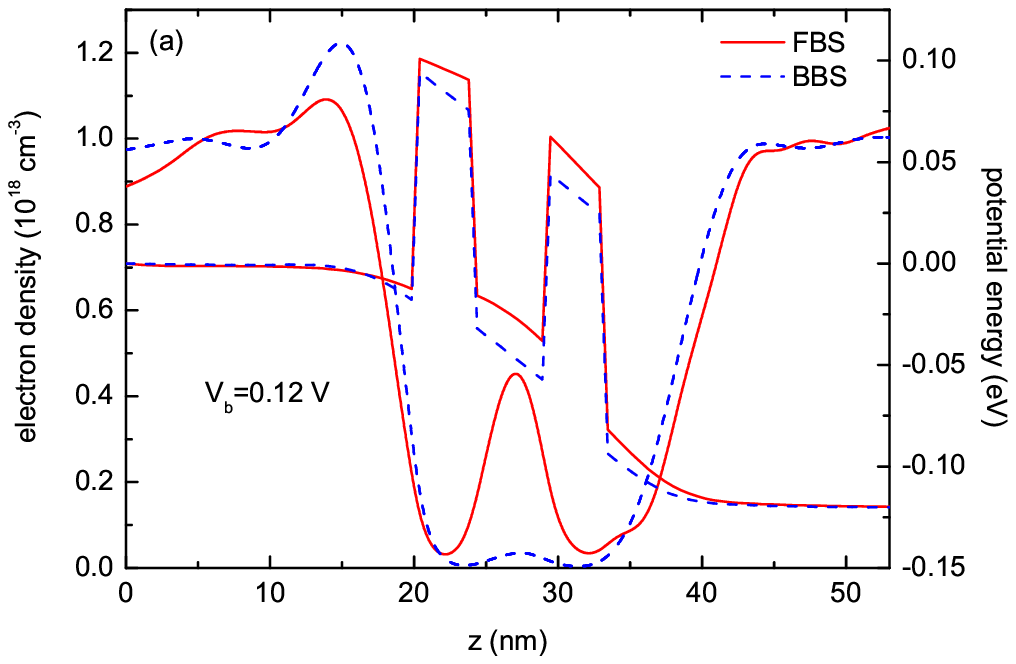}\\
\includegraphics[width=\columnwidth]{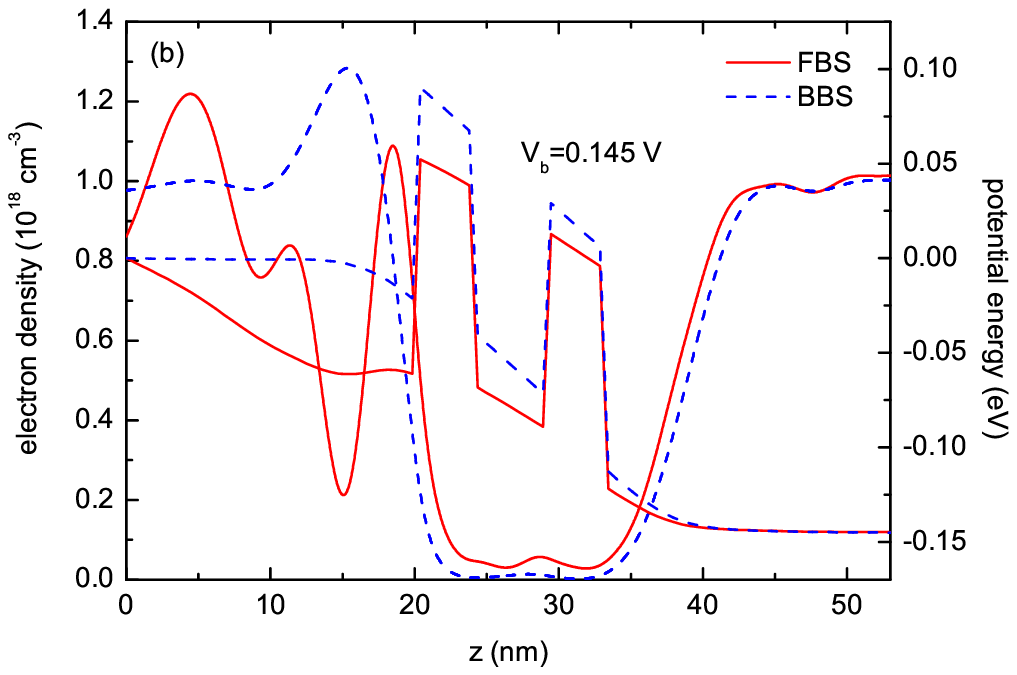}
\caption{\label{fig:7}(Color online) Electron density distributions and 
potential energy profiles for spin-up electrons at magnetic field $B=6$
T for the bias (a) $V_b=0.12$ V corresponding to range (I),
(b)  $V_b=0.145$~V corresponding to range (II) (see text).}
\end{figure}

In order to get a more deep physical insight into the origin of the 
current hysteresis we consider the current-voltage characteristics for
spin-up component of the current at the magnetic field $B=6$T for which
the electron density distributions and potential energy profiles are
calculated [Fig.~\ref{fig:7}(a,b)].
First, we discuss the effect of the current hysteresis in the first bias
voltage range below the resonant current peak.
We see [Fig. \ref{fig:7}(a)] that the potential energy profiles and the 
electron density distributions obtained for the increasing and
decreasing bias voltage are different. 
In the case of FBS, the electrons are accumulated in the paramagnetic
quantum well, which is  in contrast to the case of BBS, for which the
electron density in the quantum well is low.
Now, we consider the processes that occur if the bias voltage is
gradually changed.
The increasing bias voltage (FBS case) leads to the lowering of the
potential well energy, which causes that the energy of the quasi-bound
state in the paramagnetic quantum well also becomes lower.
As a result at the certain bias voltage, the resonant tunneling
condition is reached.
The resonant tunneling appears if the energy of the quantum-well
quasi-bound state falls into the transport window, i.e. it takes the
value between the minimum of the conduction band for $n$-ZnSe
and the electrochemical potential of the left contact.
In the resonance tunneling regime, the current density reaches the
maximum and we observe the accumulation of the electrons in the
central quantum-well region [Fig.~\ref{fig:7}(a)].  
The electrons accumulated in the quantum well create the electric field
that shifts up the potential well bottom and consequently the energy of
the quasi-bound state.  
Therefore, the effect of the further increase of the bias voltage that
leads to getting out the resonance is partly compensated by the
accumulation of electrons in the quantum well.
This causes that -- in the case of FBS -- the resonant tunneling 
condition is satisfied in a wider bias voltage range.
As might be expected this effect is not observed for the BBS case
because the decreasing bias voltage and the charge accumulated in the
quantum well shifts up the potential well bottom, which increases the
energy of the quasi-bound state in the quantum well region.
Therefore, the different effects of the increasing and decreasing bias
voltage on the quantum transport conditions allow us to explain
the appearance of the current hysteresis in bias voltage range (I).

\begin{figure}
 \includegraphics[width=\columnwidth]{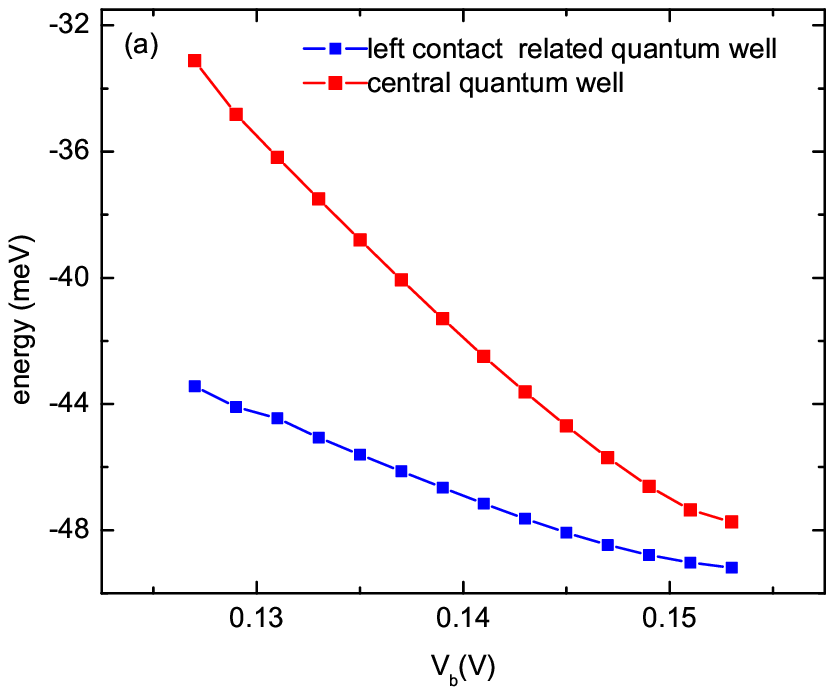}\\
 \includegraphics[width=\columnwidth]{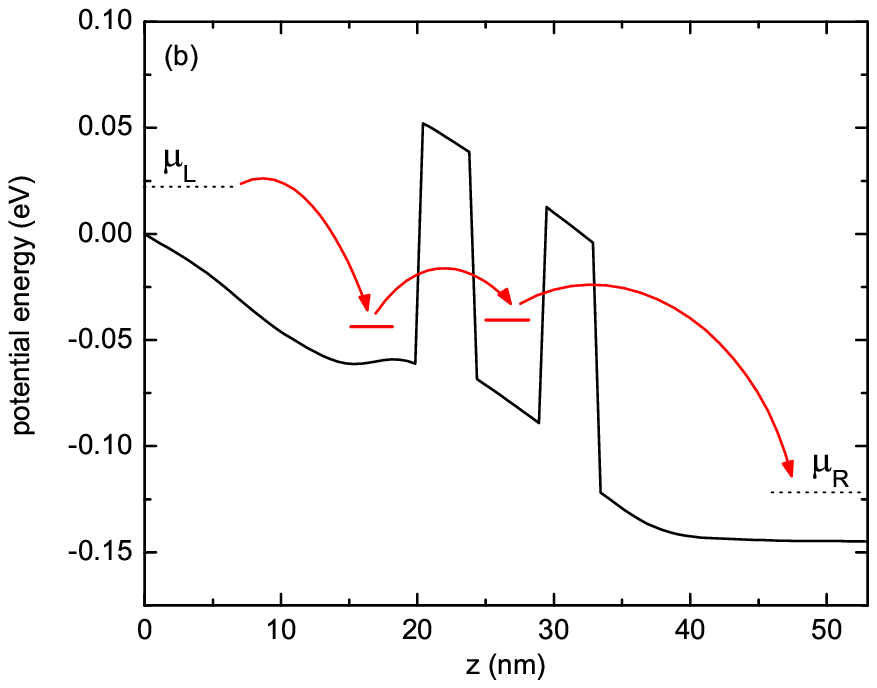}
 \caption{\label{fig:8}(a) Expectation value of the energy of the 
quasi--bound state localized in the central quantum well and in the
quantum well formed in the left contact calculated for the bias voltage
from region (II) [Fig. 2(c)].
(b) Schematic illustration of the resonant tunneling through the 
quasi-bound states of the left contact potential well and central
potential well.  
This process is responsible for the current hysteresis in bias voltage
range (II) above the resonant current peak.}
\end{figure}

The current hysteresis in bias voltage range (II) located above the bias
corresponding to the resonant current peak manifests itself as the
increase of the current in the NDR region on the current-voltage
characteristics [cf. Fig.~\ref{fig:2}].
In this range of bias voltage, the potential energy profiles and
electron density distributions obtained for the increasing and
decreasing bias voltage are different (Fig.~\ref{fig:7}(b)), they also
exhibit an essential difference in comparison to those obtained for
range (I) [cf. Fig.~\ref{fig:7}(a)].
In the case of FBS, we observe the depletion of the electron density in
the region of the left contact and the formation of the shallow
potential well in this region.
In the NDR region the system rapidly escapes from the resonant
tunneling condition and consequently the reflection probability from
the left barrier drastically increases.
The wave function of the electron incoming from the left reservoir 
interfere with the wave function of the electron reflected from the left
barrier, which leads to the electron density oscillations in the region
of the left contact [cf. Fig. \ref{fig:7}(b)].
In a self-consistent manner, this electron density distribution leads to 
creation of the shallow potential well, in which the quasi-bound
state can be formed.
The results of calculations of the energy expectation values for the
quasi-bound states in the central quantum well and in the quantum well
located in the left contact [Fig.~\ref{fig:8}(a)] allow to explain the
relation between these quasi-bound states.
It appears that for each bias voltage value from region (II) the energy
of the quasi-bound state localized in the region of the left contact
is lower than the energy of the quasi-bound state localized in the 
central quantum well.
If the bias voltage increases, the difference between these energies
decreases from $\sim 10$ meV for $V_b=0.125$~V to $\sim 2$ meV for
$V_b=0.153$~V. 
We note that this energy difference is small and comparable to the
energy broadening of the corresponding energy level.  
Therefore, the resonant tunneling conditions are satisfied for some
range of the bias voltage and the electrons can tunnel via the
quasi-bound states from the left contact through the central quantum
well to the right contact leading to the linear increase of the
current in the NDR region.
The resonant tunneling that occurs for the FBS in region (II) is
schematically illustrated in Fig.~\ref{fig:8}(b).
In the case of BBS,  the potential well in the left contact region is 
not deep enough in order to the quasi-bound state to be formed  [cf.
Fig. \ref{fig:7}(b)].
Therefore, the tunneling can appear with a very small probability and 
the BBS current is very small.
These effects lead to the current hysteresis in bias voltage range
(II) located above the resonant current peak.

\begin{figure}
\includegraphics[width=\columnwidth]{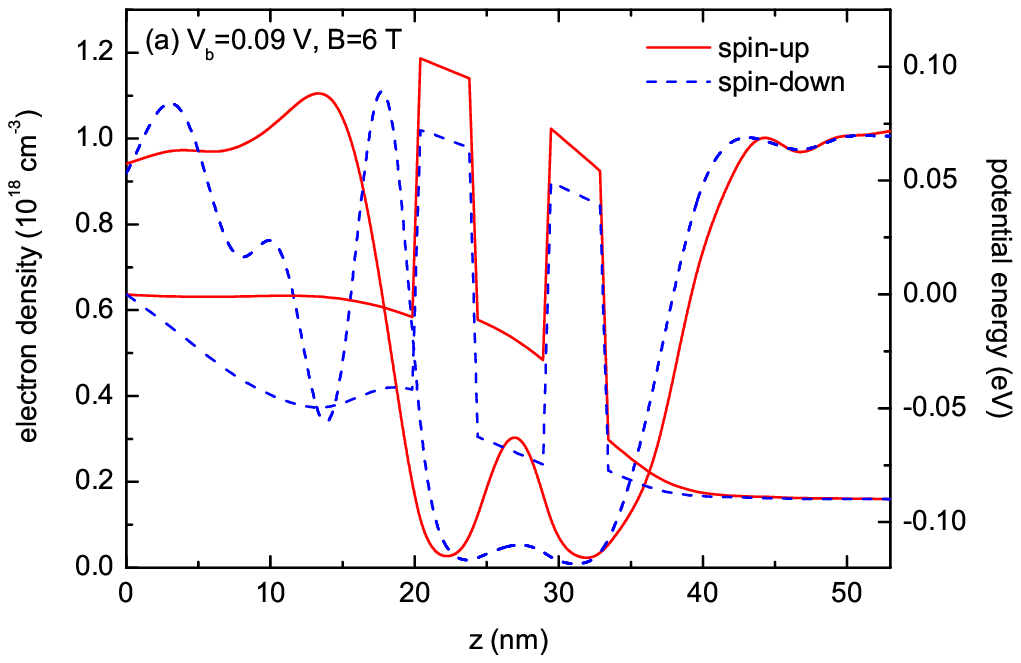}
\includegraphics[width=\columnwidth]{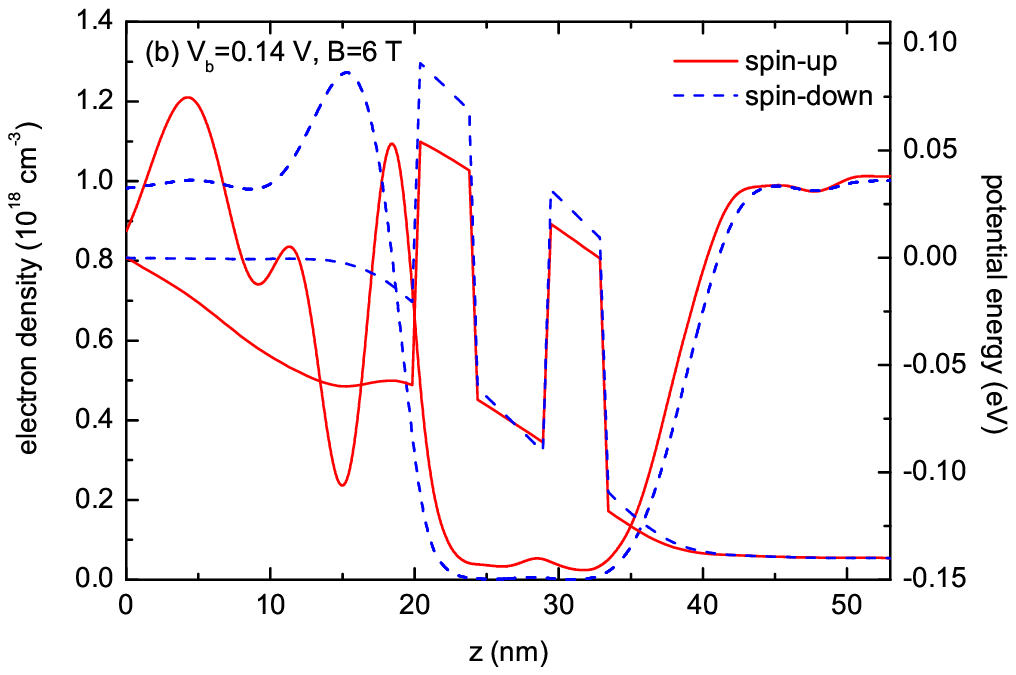}
\caption{\label{fig:9}(Color online) Electron density distributions and 
potential energy profiles for spin-up and spin-down electrons in the
case of FBS at magnetic field $B=6$~T for  bias voltage (a)
$V_b=0.009$~V and (b) $V_b=0.14$~V.}
\end{figure}

\begin{figure}
\includegraphics[width=\columnwidth]{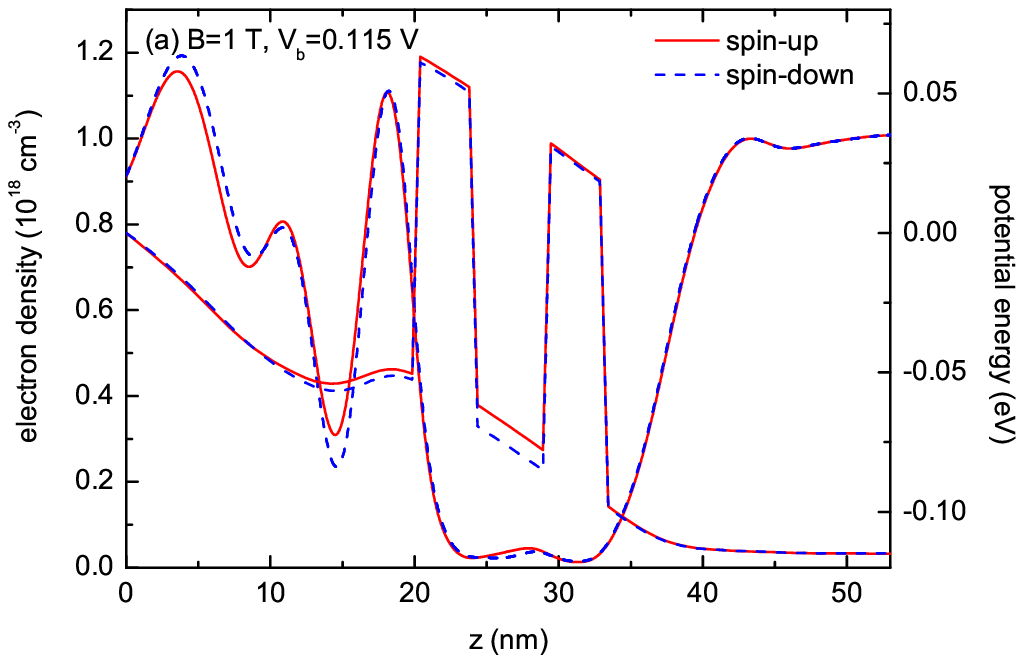}\\
\includegraphics[width=\columnwidth]{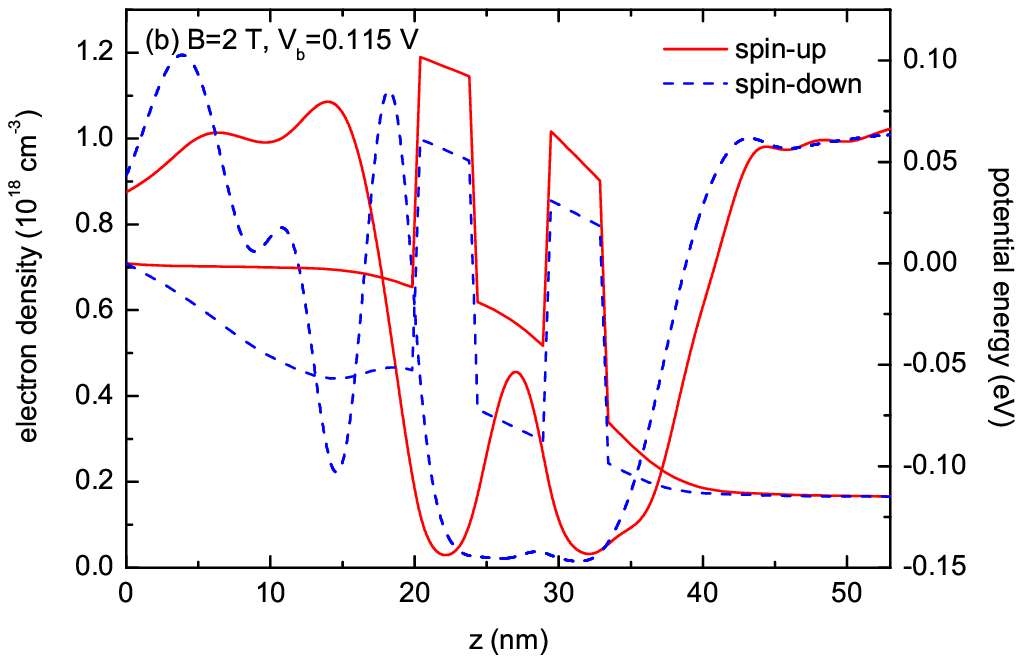}\\
\includegraphics[width=\columnwidth]{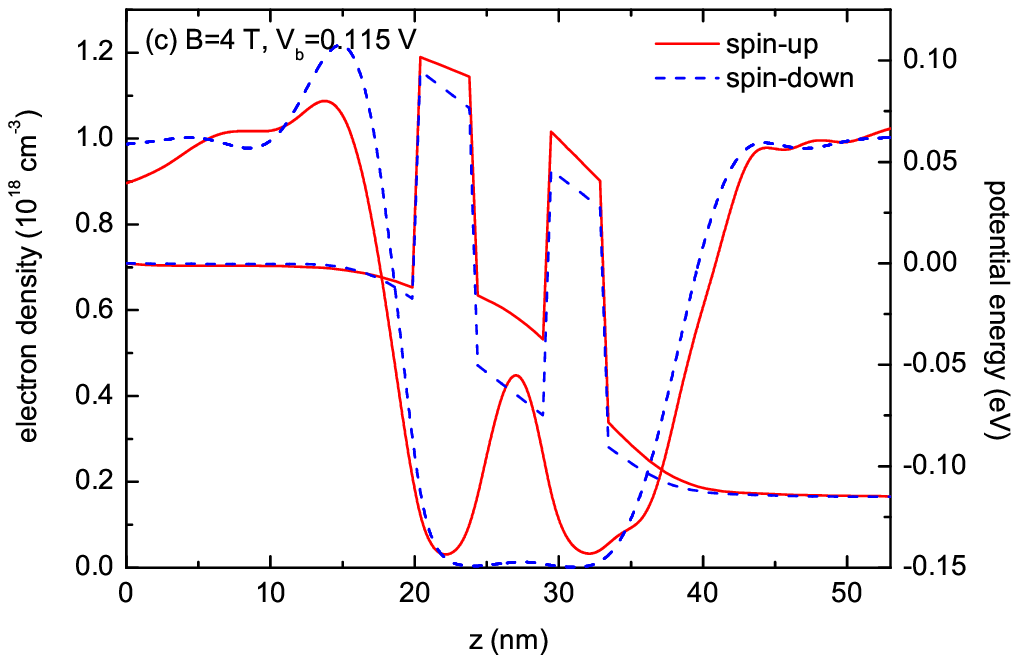}
\caption{\label{fig:10}(Color online) Electron density distributions 
and potential energy profiles for spin-up and spin-down electrons in the
case of FBS for bias voltage $V_b=0.115$~V and magnetic fields (a)
$B=1$~T, (b) $B=2$~T, and (c) $B=4$~T, which correspond to the first,
second, and third plateau of the spin polarization displayed on Fig. 6.}
\end{figure}

The plateau behavior of the spin current polarization that 
appears in certain ranges of the bias voltage (Fig. \ref{fig:5})
and magnetic field (Fig. \ref{fig:6}) are the most interesting results 
of the present calculations.
We demonstrate that all of these plateaus are caused by the tunneling of
spin-polarized electrons through the quasi-bound state in the region of
the left contact. As we mentioned before the same process leads to the
increase of the current in the NDR region and consequently the current
hysteresis occurring in bias voltage range (II).
For this purpose we analyze the spin-dependent electron distributions 
and potential energy profiles calculated for different bias voltage
values at the fixed magnetic field (Fig. \ref{fig:9})
and for different values of the magnetic fields at the fixed bias
voltage (Fig. \ref{fig:10}).
Fig.~\ref{fig:9} displays the results for the two bias voltage values
that correspond to the two plateaus of spin current polarization in the
lower and higher bias voltage regions (cf. Fig.~\ref{fig:5}).
Fig.~\ref{fig:9}(a) shows that -- for the lower bias voltage and for
the spin-down electrons -- we deal with the charge depletion in the
central quantum well and the formation of the quantum well in the region
of the left contact.
In this quantum well, the spin-down electrons can be accumulated in the
weakly bound states.
In the higher bias voltage region [Fig.~\ref{fig:9}(b)], the roles
of electron spin states are interchanged.
As a result, the charge depletion appears for the spin-up states in the 
central quantum well and the spin-up electrons become accumulated in the
left-contact related potential well.
Therefore, in both the lower and higher bias voltage regions,  the
almost constant spin current polarization results from the resonant
tunneling of the spin-polarized electrons via the quasi-bound state
formed in the region of the left contact.
The current of the spin-down electrons is responsible for the plateau 
in the lower bias voltage range, while the plateau of the spin current
polarization in the higher bias voltage range is due to the resonant
tunneling of the spin-up electrons.

The spin current polarization versus magnetic field dependence exhibits 
three plateaus for the bias voltage $V_b=0.115$~V (cf.
Fig.~\ref{fig:6}). 
In order to explain the origin of this interesting behavior we have
calculated the spin-dependent electron distributions and potential
energy profiles for magnetic fields $B$ =1~T, 2~T, and 4~T that
correspond to the first, second, and third plateau, respectively, on the
$P_j(B)$ curve (Fig.~\ref{fig:10}).
In the magnetic field range corresponding to the first plateau 
(Fig.~\ref{fig:10}(a)), we have obtained the depletion of both the
spin-up and spin-down electrons in the central quantum-well region.
Moreover, in the region of the left contact, the shallow quantum well 
is formed for both the spin-up and spin-down electrons.
In this left-contact related potential well, the quasi-bound states of 
the electrons with either spin can be created.
These quasi-bound states participate in the resonant tunneling through 
the paramagnetic RTD by means of the same process as observed in case of
the electron tunneling in the range of the current hysteresis marked
(II) on Fig.~\ref{fig:2}(c).

Since the potential energy profiles and electron densities for the 
spin-up and spin-down electrons are nearly the same, both the spin
currents are almost equal to each other, which leads to the vanishing
spin current polarization [cf. Fig. \ref{fig:10}(a)].
The increasing magnetic field changes -- in different manner -- the 
resonant tunneling conditions for each spin polarization of the current.
If magnetic field exceeds $B=1.6$~T, the quantum well for spin-up 
electrons in the region of the left contact disappears [Fig.
\ref{fig:10}(b)].
At high magnetic fields, the spin-up electrons become accumulated in 
the central quantum-well layer.
This means that in the region of the central quantum well the resonant
tunneling condition is satisfied only for the spin-up electrons,
whereas the spin-down electrons can tunnel through the quasi-bound state
in the region of left contact.
Therefore, the spin-up current component becomes large, while the 
spin-down component is only slightly changed (cf. Fig.~\ref{fig:2}).
As a result of these changes, we obtain -- at the intermediate magnetic 
field (Fig. \ref{fig:6})  -- the second plateau of the spin current
polarization with $P_j \simeq 0.5$.
If the magnetic field exceeds $\sim 3$ T, the left-contact related 
quantum well also disappears for the spin-down electrons.
At high magnetic fields, the density of the spin-up (spin-down) 
electrons is large (small) in the central quantum well.
In this case, the resonant tunneling condition is satisfied for the 
spin-up electrons, while the spin-down electrons are out of the
resonance.
Therefore, the spin-up current reaches the maximum while the spin-down 
current becomes negligibly small [cf. Fig. \ref{fig:2}(d)].
This leads to the third plateau on the $P_j(B)$ curve with 
$P_j \simeq 1$, i.e., we obtain the almost full spin current
polarization in the high magnetic field ranges (Fig. \ref{fig:6}).
We note that for the lower and higher bias voltage the system makes the
direct transition to the fully polarized spin current at certain
magnetic field (cf. Fig.~\ref{fig:6}).

\section{Conclusions and Summary}\label{sec:concl}

In the present paper, we have studied the spin-dependent electron 
transport through the paramagnetic RTD based on
Zn$_{1-x}$Mn$_x$Se/Zn$_{0.95}$Be$_{0.05}$Se heterostructure.
The current-voltage characteristics of the nanodevice have been
determined by the numerical solution of self-consistent Wigner-Poisson
model of electron transport in the framework of the two-current model.
The application of the two-current model is fully justified in the case
of the nanodevice because the magnetic field produces the giant Zeeman
splitting [Eq.~(\ref{eq:GZS})] which modifies the potential energy
profile and as a result the electrons with opposite spins
independently tunnel via the different quantum-well states.
Numerical estimation of the ratio of the cyclotron resonance energy
$\hbar \omega_c$, where $\omega_c = eB/m_e$, to the Fermi energy of
$n$-doped ZnSe for maximum value of magnetic field ($B$ = 8 T) which
is considered in this paper gives $\sim$0.46.
It means that the relative strength of the maximum magnetic field is 
rather weak and therefore we can neglect the effect of magnetic field
quantization.
Due to the complexity of the numerical calculations, we have neglected
the  intervalley coupling and effects of the electron scattering
processes which result from the interaction of electrons with
phonons and impurities.

The calculated current-voltage characteristics exhibit two separated 
resonant current peaks that correspond to the spin-up and spin-down
current components.
We have found that the current hysteresis appears on the
current-voltage characteristics for each spin polarization of the
current.
The current hysteresis occurs in two distinct bias voltage regions:
the first with the bias voltage values below that corresponding to the
resonant current peak and the second -- above the resonant current peak.
The analysis of the self-consistent potential energy profiles and 
electron density distributions in the nanodevice allows us to give the
physical interpretation of these effects.
We have demonstrated that the current hysteresis in the first bias
voltage region is related to the charge accumulation in the central
quantum-well layer, while the current hysteresis in the second bias
voltage region results from the resonant tunneling through the
quasi-bound state created in the region of the left contact.
We have shown that the current hysteresis leads to the different spin 
polarization for the forward and backward bias sweeps,
which -- in turn -- leads to the appearance of the plateaus that appear 
on the spin current polarization dependence on the bias voltage and
magnetic field.  
For certain plateaus the current is almost fully spin-polarized.
We have determined the conditions for the full spin current
polarization and discussed the underlying physics.
This property is very promising for future spintronic devices.

We expect that including the scattering of electrons should result in 
more smooth current-voltage characteristics and the decrease of
the spin current polarization.
Moreover, the exchange interaction between the conduction-band 
electrons has been neglected in the present calculations.
We have also performed the calculations with the exchange interaction 
taken into account by the local density approximation.  
The preliminary results show that the incorporation of the interelectron
exchange changes the present results only slightly, in particular, the
maximum current polarization becomes slightly smaller and the rapid
changes on the current-voltage characteristics become more smooth.  
We conclude that the exchange interaction between the conduction-band
electrons has no significant influence on the present results.

In summary, we have shown that the electron current flowing through the 
non-magnetic$/$magnetic$/$non-magnetic double barrier heterostructure
(paramagnetic RTD) can exhibit the hysteresis that appears on the
current-voltage characteristics for the spin-polarized current.  
Moreover, we have determined the regions of the bias voltage and ranges
of magnetic field, in which the pronounced spin current polarization
occurs in the paramagnetic RTD.
The results of the present paper allow us to predict the conditions, 
under which we obtain the spin-polarized current in the double-barrier
heterostructure with the manganese doped quantum-well layer.

\begin{acknowledgments}
This paper has been supported by the Foundation for Polish Science MPD 
Programme co-financed by the EU European Regional Development Fund
and  by the AGH-University of Science and Technology under project 
no. 11.11.220.01 ''Basic and applied  research in nuclear and solid
state physics''.
\end{acknowledgments}


\begin{thebibliography}{10}%
\makeatletter
\providecommand \@ifxundefined [1]{%
 \ifx #1\undefined \expandafter \@firstoftwo
 \else \expandafter \@secondoftwo
\fi
}%
\providecommand \@ifnum [1]{%
 \ifnum #1\expandafter \@firstoftwo
 \else \expandafter \@secondoftwo
\fi
}%
\providecommand \enquote [1]{``#1''}%
\providecommand \bibnamefont  [1]{#1}%
\providecommand \bibfnamefont [1]{#1}%
\providecommand \citenamefont [1]{#1}%
\providecommand\href[0]{\@sanitize\@href}%
\providecommand\@href[1]{\endgroup\@@startlink{#1}\endgroup\@@href}%
\providecommand\@@href[1]{#1\@@endlink}%
\providecommand \@sanitize [0]{\begingroup\catcode`\&12\catcode`\#12\relax}%
\@ifxundefined \pdfoutput {\@firstoftwo}{%
 \@ifnum{\z@=\pdfoutput}{\@firstoftwo}{\@secondoftwo}%
}{%
 \providecommand\@@startlink[1]{\leavevmode}%
 \providecommand\@@endlink[0]{}%
}{%
 \providecommand\@@startlink[1]{%
  \leavevmode
  \pdfstartlink
   attr{/Border[0 0 1 ]/H/I/C[0 1 1]}%
   user{/Subtype/Link/A<</Type/Action/S/URI/URI(#1)>>}%
  \relax
 }%
 \providecommand\@@endlink[0]{\pdfendlink}%
}%
\providecommand \url  [0]{\begingroup\@sanitize \@url }%
\providecommand \@url [1]{\endgroup\@href {#1}{\urlprefix}}%
\providecommand \urlprefix [0]{URL }%
\providecommand \Eprint[0]{\href }%
\@ifxundefined \urlstyle {%
  \providecommand \doi [1]{doi:\discretionary{}{}{}#1}%
}{%
  \providecommand \doi [0]{doi:\discretionary{}{}{}\begingroup
  \urlstyle{rm}\Url }%
}%
\providecommand \doibase [0]{http://dx.doi.org/}%
\providecommand \Doi[1]{\href{\doibase#1}}%
\providecommand \bibAnnote [3]{%
  \BibitemShut{#1}%
  \begin{quotation}\noindent
    \textsc{Key:}\ #2\\\textsc{Annotation:}\ #3%
  \end{quotation}%
}%
\providecommand \bibAnnoteFile [2]{%
  \IfFileExists{#2}{\bibAnnote {#1} {#2} {\input{#2}}}{}%
}%
\providecommand \typeout [0]{\immediate \write \m@ne }%
\providecommand \selectlanguage [0]{\@gobble}%
\providecommand \bibinfo [0]{\@secondoftwo}%
\providecommand \bibfield [0]{\@secondoftwo}%
\providecommand \translation [1]{[#1]}%
\providecommand \BibitemOpen[0]{}%
\providecommand \bibitemStop [0]{}%
\providecommand \bibitemNoStop [0]{.\EOS\space}%
\providecommand \EOS [0]{\spacefactor3000\relax}%
\providecommand \BibitemShut [1]{\csname bibitem#1\endcsname}%
\bibitem{Ohno1992}%
  \BibitemOpen
  \bibfield{author}{%
  \bibinfo {author} {\bibfnamefont{H.}~\bibnamefont{Ohno}}, \bibinfo {author}
  {\bibfnamefont{H.}~\bibnamefont{Munekata}}, \bibinfo {author}
  {\bibfnamefont{T.}~\bibnamefont{Penney}}, \bibinfo {author}
  {\bibfnamefont{S.}~\bibnamefont{von Moln{\'a}r}},\ and\ \bibinfo {author}
  {\bibfnamefont{L.~L.}\ \bibnamefont{Chang}},\ }%
  \bibfield{journal}{%
  \bibinfo {journal} {Phys. Rev. Lett.}\ }%
  \textbf{\bibinfo {volume} {68}},\ \bibinfo {pages} {2664} (\bibinfo {year}
  {1992})%
  \bibAnnoteFile{NoStop}{Ohno1992}%
\bibitem{Ohno1998}%
  \BibitemOpen
  \bibfield{author}{%
  \bibinfo {author} {\bibfnamefont{H.}~\bibnamefont{Ohno}},\ }%
  \bibfield{journal}{%
  \bibinfo {journal} {Science}\ }%
  \textbf{\bibinfo {volume} {281}},\ \bibinfo {pages} {951} (\bibinfo {year}
  {1998})%
  \bibAnnoteFile{NoStop}{Ohno1998}%
\bibitem{Hayashi2001}%
  \BibitemOpen
  \bibfield{author}{%
  \bibinfo {author} {\bibfnamefont{T.}~\bibnamefont{Hayashi}}, \bibinfo
  {author} {\bibfnamefont{Y.}~\bibnamefont{Hashimoto}}, \bibinfo {author}
  {\bibfnamefont{S.}~\bibnamefont{Katsumoto}},\ and\ \bibinfo {author}
  {\bibfnamefont{Y.}~\bibnamefont{Iye}},\ }%
  \bibfield{journal}{%
  \bibinfo {journal} {Appl. Phys. Lett.}\ }%
  \textbf{\bibinfo {volume} {78}},\ \bibinfo {pages} {1691} (\bibinfo {year}
  {2001})%
  \bibAnnoteFile{NoStop}{Hayashi2001}%
\bibitem{Edmonds2002}%
  \BibitemOpen
  \bibfield{author}{%
  \bibinfo {author} {\bibfnamefont{K.~W.}\ \bibnamefont{Edmonds}}, \bibinfo
  {author} {\bibfnamefont{K.~Y.}\ \bibnamefont{Wang}}, \bibinfo {author}
  {\bibfnamefont{R.~P.}\ \bibnamefont{Campion}}, \bibinfo {author}
  {\bibfnamefont{A.~C.}\ \bibnamefont{Neumann}}, \bibinfo {author}
  {\bibfnamefont{N.~R.~S.}\ \bibnamefont{Farley}}, \bibinfo {author}
  {\bibfnamefont{B.~L.}\ \bibnamefont{Gallagher}},\ and\ \bibinfo {author}
  {\bibfnamefont{C.~T.}\ \bibnamefont{Foxon}},\ }%
  \bibfield{journal}{%
  \bibinfo {journal} {Appl. Phys. Lett.}\ }%
  \textbf{\bibinfo {volume} {81}},\ \bibinfo {pages} {4991} (\bibinfo {year}
  {2002})%
  \bibAnnoteFile{NoStop}{Edmonds2002}%
\bibitem{Yu2002}%
  \BibitemOpen
  \bibfield{author}{%
  \bibinfo {author} {\bibfnamefont{K.~M.}\ \bibnamefont{Yu}}, \bibinfo {author}
  {\bibfnamefont{W.}~\bibnamefont{Walukiewicz}}, \bibinfo {author}
  {\bibfnamefont{T.}~\bibnamefont{Wojtowicz}}, \bibinfo {author}
  {\bibfnamefont{I.}~\bibnamefont{Kuryliszyn}}, \bibinfo {author}
  {\bibfnamefont{X.}~\bibnamefont{Liu}}, \bibinfo {author}
  {\bibfnamefont{Y.}~\bibnamefont{Sasaki}},\ and\ \bibinfo {author}
  {\bibfnamefont{J.~K.}\ \bibnamefont{Furdyna}},\ }%
  \bibfield{journal}{%
  \bibinfo {journal} {Phys. Rev. B}\ }%
  \textbf{\bibinfo {volume} {65}},\ \bibinfo {pages} {201303} (\bibinfo {year}
  {2002})%
  \bibAnnoteFile{NoStop}{Yu2002}%
\bibitem{Chiba2003}%
  \BibitemOpen
  \bibfield{author}{%
  \bibinfo {author} {\bibfnamefont{D.}~\bibnamefont{Chiba}}, \bibinfo {author}
  {\bibfnamefont{K.}~\bibnamefont{Takamura}}, \bibinfo {author}
  {\bibfnamefont{F.}~\bibnamefont{Matsukura}},\ and\ \bibinfo {author}
  {\bibfnamefont{H.}~\bibnamefont{Ohno}},\ }%
  \bibfield{journal}{%
  \bibinfo {journal} {Appl. Phys. Lett.}\ }%
  \textbf{\bibinfo {volume} {82}},\ \bibinfo {pages} {3020} (\bibinfo {year}
  {2003})%
  \bibAnnoteFile{NoStop}{Chiba2003}%
\bibitem{Ku2003}%
  \BibitemOpen
  \bibfield{author}{%
  \bibinfo {author} {\bibfnamefont{K.~C.}\ \bibnamefont{Ku}}, \bibinfo {author}
  {\bibfnamefont{S.~J.}\ \bibnamefont{Potashnik}}, \bibinfo {author}
  {\bibfnamefont{R.~F.}\ \bibnamefont{Wang}}, \bibinfo {author}
  {\bibfnamefont{S.~H.}\ \bibnamefont{Chun}}, \bibinfo {author}
  {\bibfnamefont{P.}~\bibnamefont{Schiffer}}, \bibinfo {author}
  {\bibfnamefont{N.}~\bibnamefont{Samarth}}, \bibinfo {author}
  {\bibfnamefont{S.~M.}\ \bibnamefont{J}}, \bibinfo {author}
  {\bibfnamefont{A.}~\bibnamefont{Mascarenhas}}, \bibinfo {author}
  {\bibfnamefont{E.}~\bibnamefont{Johnston-Halperin}}, \bibinfo {author}
  {\bibfnamefont{R.~C.}\ \bibnamefont{Myers}}, \bibinfo {author}
  {\bibfnamefont{A.~C.}\ \bibnamefont{Gossard}},\ and\ \bibinfo {author}
  {\bibfnamefont{D.~D.}\ \bibnamefont{Awschalom}},\ }%
  \bibfield{journal}{%
  \bibinfo {journal} {Appl. Phys. Lett.}\ }%
  \textbf{\bibinfo {volume} {82}},\ \bibinfo {pages} {2302} (\bibinfo {year}
  {2003})%
  \bibAnnoteFile{NoStop}{Ku2003}%
\bibitem{Eid2005}%
  \BibitemOpen
  \bibfield{author}{%
  \bibinfo {author} {\bibfnamefont{K.~F.}\ \bibnamefont{Eid}}, \bibinfo
  {author} {\bibfnamefont{B.~L.}\ \bibnamefont{Sheu}}, \bibinfo {author}
  {\bibfnamefont{O.}~\bibnamefont{Maksimov}}, \bibinfo {author}
  {\bibfnamefont{M.~B.}\ \bibnamefont{Stone}}, \bibinfo {author}
  {\bibfnamefont{P.}~\bibnamefont{Schiffer}},\ and\ \bibinfo {author}
  {\bibfnamefont{N.}~\bibnamefont{Samarth}},\ }%
  \bibfield{journal}{%
  \bibinfo {journal} {Appl. Phys. Lett.}\ }%
  \textbf{\bibinfo {volume} {86}},\ \bibinfo {pages} {152505} (\bibinfo {year}
  {2005})%
  \bibAnnoteFile{NoStop}{Eid2005}%
\bibitem{Jungwirth2005}%
  \BibitemOpen
  \bibfield{author}{%
  \bibinfo {author} {\bibfnamefont{T.}~\bibnamefont{Jungwirth}}, \bibinfo
  {author} {\bibfnamefont{K.~Y.}\ \bibnamefont{Wang}}, \bibinfo {author}
  {\bibfnamefont{J.}~\bibnamefont{Masek}}, \bibinfo {author}
  {\bibfnamefont{K.~W.}\ \bibnamefont{Edmonds}}, \bibinfo {author}
  {\bibfnamefont{J.}~\bibnamefont{K{\"o}nig}}, \bibinfo {author}
  {\bibfnamefont{J.}~\bibnamefont{Sinova}}, \bibinfo {author}
  {\bibfnamefont{M.}~\bibnamefont{Polini}}, \bibinfo {author}
  {\bibfnamefont{N.~A.}\ \bibnamefont{Goncharuk}}, \bibinfo {author}
  {\bibfnamefont{A.~H.}\ \bibnamefont{MacDonald}}, \bibinfo {author}
  {\bibfnamefont{M.}~\bibnamefont{Sawicki}}, \bibinfo {author}
  {\bibfnamefont{R.~P.}\ \bibnamefont{Campion}}, \bibinfo {author}
  {\bibfnamefont{L.~X.}\ \bibnamefont{Zhao}}, \bibinfo {author}
  {\bibfnamefont{C.~T.}\ \bibnamefont{Foxon}},\ and\ \bibinfo {author}
  {\bibfnamefont{B.~L.}\ \bibnamefont{Gallagher}},\ }%
  \bibfield{journal}{%
  \bibinfo {journal} {Phys. Rev. B}\ }%
  \textbf{\bibinfo {volume} {72}},\ \bibinfo {pages} {165204} (\bibinfo {year}
  {2005})%
  \bibAnnoteFile{NoStop}{Jungwirth2005}%
\bibitem{Furdyna1988}%
  \BibitemOpen
  \bibfield{author}{%
  \bibinfo {author} {\bibfnamefont{J.~K.}\ \bibnamefont{Furdyna}},\ }%
  \bibfield{journal}{%
  \bibinfo {journal} {J. Appl. Phys.}\ }%
  \textbf{\bibinfo {volume} {64}},\ \bibinfo {pages} {R29} (\bibinfo {year}
  {1988})%
  \bibAnnoteFile{NoStop}{Furdyna1988}%
\bibitem{Yu1995}%
  \BibitemOpen
  \bibfield{author}{%
  \bibinfo {author} {\bibfnamefont{W.~Y.}\ \bibnamefont{Yu}}, \bibinfo {author}
  {\bibfnamefont{A.}~\bibnamefont{Twardowski}}, \bibinfo {author}
  {\bibfnamefont{L.~P.}\ \bibnamefont{Fu}},\ and\ \bibinfo {author}
  {\bibfnamefont{A.}~\bibnamefont{Petrou}},\ }%
  \bibfield{journal}{%
  \bibinfo {journal} {Phys. Rev. B}\ }%
  \textbf{\bibinfo {volume} {51}},\ \bibinfo {pages} {9722} (\bibinfo {year}
  {1995})%
  \bibAnnoteFile{NoStop}{Yu1995}%
\bibitem{Ando2003}%
  \BibitemOpen
  \bibfield{author}{%
  \bibinfo {author} {\bibfnamefont{K.}~\bibnamefont{Ando}},\ }%
  \bibfield{journal}{%
  \bibinfo {journal} {Appl. Phys. Lett.}\ }%
  \textbf{\bibinfo {volume} {82}},\ \bibinfo {pages} {100} (\bibinfo {year}
  {2003})%
  \bibAnnoteFile{NoStop}{Ando2003}%
\bibitem{Liu2005}%
  \BibitemOpen
  \bibfield{author}{%
  \bibinfo {author} {\bibfnamefont{C.}~\bibnamefont{Liu}}, \bibinfo {author}
  {\bibfnamefont{F.}~\bibnamefont{Yun}},\ and\ \bibinfo {author}
  {\bibfnamefont{H.}~\bibnamefont{Morkor}},\ }%
  \bibfield{journal}{%
  \bibinfo {journal} {J. Mater. Sci.: Mater. Electron.}\ }%
  \textbf{\bibinfo {volume} {16}},\ \bibinfo {pages} {555} (\bibinfo {year}
  {2005})%
  \bibAnnoteFile{NoStop}{Liu2005}%
\bibitem{Jungwirth2006}%
  \BibitemOpen
  \bibfield{author}{%
  \bibinfo {author} {\bibfnamefont{T.}~\bibnamefont{Jungwirth}}, \bibinfo
  {author} {\bibfnamefont{J.}~\bibnamefont{Sinova}}, \bibinfo {author}
  {\bibfnamefont{J.}~\bibnamefont{Masek}}, \bibinfo {author}
  {\bibfnamefont{J.}~\bibnamefont{Kucera}},\ and\ \bibinfo {author}
  {\bibfnamefont{A.~H.}\ \bibnamefont{MacDonald}},\ }%
  \bibfield{journal}{%
  \bibinfo {journal} {Rev. Mod. Phys.}\ }%
  \textbf{\bibinfo {volume} {78}},\ \bibinfo {pages} {809} (\bibinfo {year}
  {2006})%
  \bibAnnoteFile{NoStop}{Jungwirth2006}%
\bibitem{Dietl2001}%
  \BibitemOpen
  \bibfield{author}{%
  \bibinfo {author} {\bibfnamefont{T.}~\bibnamefont{Dietl}}, \bibinfo {author}
  {\bibfnamefont{H.}~\bibnamefont{Ohno}},\ and\ \bibinfo {author}
  {\bibfnamefont{F.}~\bibnamefont{Matsukura}},\ }%
  \bibfield{journal}{%
  \bibinfo {journal} {Phys. Rev. B}\ }%
  \textbf{\bibinfo {volume} {63}},\ \bibinfo {pages} {195205} (\bibinfo {year}
  {2001})%
  \bibAnnoteFile{NoStop}{Dietl2001}%
\bibitem{Gruber2001}%
  \BibitemOpen
  \bibfield{author}{%
  \bibinfo {author} {\bibfnamefont{T.}~\bibnamefont{Gruber}}, \bibinfo {author}
  {\bibfnamefont{M.}~\bibnamefont{Keim}}, \bibinfo {author}
  {\bibfnamefont{R.}~\bibnamefont{Fiederling}}, \bibinfo {author}
  {\bibfnamefont{G.}~\bibnamefont{Reuscher}}, \bibinfo {author}
  {\bibfnamefont{W.}~\bibnamefont{Ossau}}, \bibinfo {author}
  {\bibfnamefont{G.}~\bibnamefont{Schmidt}}, \bibinfo {author}
  {\bibfnamefont{L.~W.}\ \bibnamefont{Molenkamp}},\ and\ \bibinfo {author}
  {\bibfnamefont{A.}~\bibnamefont{Waag}},\ }%
  \bibfield{journal}{%
  \bibinfo {journal} {Appl. Phys. Lett.}\ }%
  \textbf{\bibinfo {volume} {78}},\ \bibinfo {pages} {1101} (\bibinfo {year}
  {2001})%
  \bibAnnoteFile{NoStop}{Gruber2001}%
\bibitem{Slobodskyy2003}%
  \BibitemOpen
  \bibfield{author}{%
  \bibinfo {author} {\bibfnamefont{A.}~\bibnamefont{Slobodskyy}}, \bibinfo
  {author} {\bibfnamefont{C.}~\bibnamefont{Gould}}, \bibinfo {author}
  {\bibfnamefont{T.}~\bibnamefont{Slobodskyy}}, \bibinfo {author}
  {\bibfnamefont{C.~R.}\ \bibnamefont{Becker}}, \bibinfo {author}
  {\bibfnamefont{G.}~\bibnamefont{Schmidt}},\ and\ \bibinfo {author}
  {\bibfnamefont{L.~W.}\ \bibnamefont{Molenkamp}},\ }%
  \bibfield{journal}{%
  \bibinfo {journal} {Phys. Rev. Lett.}\ }%
  \textbf{\bibinfo {volume} {90}},\ \bibinfo {pages} {246601} (\bibinfo {year}
  {2003})%
  \bibAnnoteFile{NoStop}{Slobodskyy2003}%
\bibitem{Beletskii2003}%
  \BibitemOpen
  \bibfield{author}{%
  \bibinfo {author} {\bibfnamefont{N.~N.}\ \bibnamefont{Beletskii}}, \bibinfo
  {author} {\bibfnamefont{G.~P.}\ \bibnamefont{Berman}},\ and\ \bibinfo
  {author} {\bibfnamefont{S.~A.}\ \bibnamefont{Borysenko}},\ }%
  \bibfield{journal}{%
  \bibinfo {journal} {Phys. Rev. B}\ }%
  \textbf{\bibinfo {volume} {71}},\ \bibinfo {pages} {125325} (\bibinfo {year}
  {2003})%
  \bibAnnoteFile{NoStop}{Beletskii2003}%
\bibitem{Datta1990}%
  \BibitemOpen
  \bibfield{author}{%
  \bibinfo {author} {\bibfnamefont{S.}~\bibnamefont{Datta}}\ and\ \bibinfo
  {author} {\bibfnamefont{B.}~\bibnamefont{Das}},\ }%
  \bibfield{journal}{%
  \bibinfo {journal} {Appl. Phys. Lett.}\ }%
  \textbf{\bibinfo {volume} {56}},\ \bibinfo {pages} {665} (\bibinfo {year}
  {1990})%
  \bibAnnoteFile{NoStop}{Datta1990}%
\bibitem{Egues1998}%
  \BibitemOpen
  \bibfield{author}{%
  \bibinfo {author} {\bibfnamefont{J.~C.}\ \bibnamefont{Egues}},\ }%
  \bibfield{journal}{%
  \bibinfo {journal} {Phys. Rev. Lett.}\ }%
  \textbf{\bibinfo {volume} {80}},\ \bibinfo {pages} {4578} (\bibinfo {year}
  {1998})%
  \bibAnnoteFile{NoStop}{Egues1998}%
\bibitem{Havu2005}%
  \BibitemOpen
  \bibfield{author}{%
  \bibinfo {author} {\bibfnamefont{P.}~\bibnamefont{Havu}}, \bibinfo {author}
  {\bibfnamefont{N.}~\bibnamefont{Tuomisto}}, \bibinfo {author}
  {\bibfnamefont{R.}~\bibnamefont{V{\"a}{\"a}n{\"a}ne}}, \bibinfo {author}
  {\bibfnamefont{M.~J.}\ \bibnamefont{Puska}},\ and\ \bibinfo {author}
  {\bibfnamefont{R.~M.}\ \bibnamefont{Nieminen}},\ }%
  \bibfield{journal}{%
  \bibinfo {journal} {Phys. Rev. B}\ }%
  \textbf{\bibinfo {volume} {71}},\ \bibinfo {pages} {235301} (\bibinfo {year}
  {2005})%
  \bibAnnoteFile{NoStop}{Havu2005}%
\bibitem{Goldman1987}%
  \BibitemOpen
  \bibfield{author}{%
  \bibinfo {author} {\bibfnamefont{V.~J.}\ \bibnamefont{Goldman}}, \bibinfo
  {author} {\bibfnamefont{D.~C.}\ \bibnamefont{Tsui}},\ and\ \bibinfo {author}
  {\bibfnamefont{J.~E.}\ \bibnamefont{Cunningham}},\ }%
  \bibfield{journal}{%
  \bibinfo {journal} {Phys. Rev. Lett.}\ }%
  \textbf{\bibinfo {volume} {58}},\ \bibinfo {pages} {1256} (\bibinfo {year}
  {1987})%
  \bibAnnoteFile{NoStop}{Goldman1987}%
\bibitem{Sheard1988}%
  \BibitemOpen
  \bibfield{author}{%
  \bibinfo {author} {\bibfnamefont{F.~W.}\ \bibnamefont{Sheard}}\ and\ \bibinfo
  {author} {\bibfnamefont{G.~A.}\ \bibnamefont{Toombs}},\ }%
  \bibfield{journal}{%
  \bibinfo {journal} {Appl. Phys. Lett.}\ }%
  \textbf{\bibinfo {volume} {52}},\ \bibinfo {pages} {1228} (\bibinfo {year}
  {1988})%
  \bibAnnoteFile{NoStop}{Sheard1988}%
\bibitem{Sollner1987}%
  \BibitemOpen
  \bibfield{author}{%
  \bibinfo {author} {\bibfnamefont{T.~C. L.~G.}\ \bibnamefont{Sollner}},\ }%
  \bibfield{journal}{%
  \bibinfo {journal} {Phys. Rev. Lett.}\ }%
  \textbf{\bibinfo {volume} {59}},\ \bibinfo {pages} {1622} (\bibinfo {year}
  {1987})%
  \bibAnnoteFile{NoStop}{Sollner1987}%
\bibitem{Su1991}%
  \BibitemOpen
  \bibfield{author}{%
  \bibinfo {author} {\bibfnamefont{B.}~\bibnamefont{Su}}, \bibinfo {author}
  {\bibfnamefont{V.~J.}\ \bibnamefont{Goldman}}, \bibinfo {author}
  {\bibfnamefont{M.}~\bibnamefont{Santos}},\ and\ \bibinfo {author}
  {\bibfnamefont{M.}~\bibnamefont{Shayegan}},\ }%
  \bibfield{journal}{%
  \bibinfo {journal} {Appl. Phys. Lett.}\ }%
  \textbf{\bibinfo {volume} {58}},\ \bibinfo {pages} {747} (\bibinfo {year}
  {1991})%
  \bibAnnoteFile{NoStop}{Su1991}%
\bibitem{Macks1996}%
  \BibitemOpen
  \bibfield{author}{%
  \bibinfo {author} {\bibfnamefont{L.~D.}\ \bibnamefont{Macks}}, \bibinfo
  {author} {\bibfnamefont{S.~A.}\ \bibnamefont{Brown}}, \bibinfo {author}
  {\bibfnamefont{R.~G.}\ \bibnamefont{Clark}}, \bibinfo {author}
  {\bibfnamefont{R.~P.}\ \bibnamefont{Starrett}}, \bibinfo {author}
  {\bibfnamefont{M.~A.}\ \bibnamefont{Reed}}, \bibinfo {author}
  {\bibfnamefont{M.~R.}\ \bibnamefont{Deshpande}}, \bibinfo {author}
  {\bibfnamefont{C.~J.~L.}\ \bibnamefont{Fernando}},\ and\ \bibinfo {author}
  {\bibfnamefont{W.~R.}\ \bibnamefont{Frensley}},\ }%
  \bibfield{journal}{%
  \bibinfo {journal} {Phys. Rev. B}\ }%
  \textbf{\bibinfo {volume} {54}},\ \bibinfo {pages} {4857} (\bibinfo {year}
  {1996})%
  \bibAnnoteFile{NoStop}{Macks1996}%
\bibitem{Goldman1987b}%
  \BibitemOpen
  \bibfield{author}{%
  \bibinfo {author} {\bibfnamefont{V.~J.}\ \bibnamefont{Goldman}}, \bibinfo
  {author} {\bibfnamefont{D.~C.}\ \bibnamefont{Tsui}},\ and\ \bibinfo {author}
  {\bibfnamefont{J.~E.}\ \bibnamefont{Cunningham}},\ }%
  \bibfield{journal}{%
  \bibinfo {journal} {Phys. Rev. B}\ }%
  \textbf{\bibinfo {volume} {36}},\ \bibinfo {pages} {7635} (\bibinfo {year}
  {1987})%
  \bibAnnoteFile{NoStop}{Goldman1987b}%
\bibitem{Dai2006}%
  \BibitemOpen
  \bibfield{author}{%
  \bibinfo {author} {\bibfnamefont{Z.}~\bibnamefont{Dai}}\ and\ \bibinfo
  {author} {\bibfnamefont{J.}~\bibnamefont{Ni}},\ }%
  \bibfield{journal}{%
  \bibinfo {journal} {Phys. Rev. B}\ }%
  \textbf{\bibinfo {volume} {73}},\ \bibinfo {pages} {113309} (\bibinfo {year}
  {2006})%
  \bibAnnoteFile{NoStop}{Dai2006}%
\bibitem{Twardowski2005}%
  \BibitemOpen
  \bibfield{author}{%
  \bibinfo {author} {\bibfnamefont{A.}~\bibnamefont{Twardowski}}, \bibinfo
  {author} {\bibfnamefont{M.}~\bibnamefont{von Ortenberg}}, \bibinfo {author}
  {\bibfnamefont{M.}~\bibnamefont{Demianiuk}},\ and\ \bibinfo {author}
  {\bibfnamefont{R.}~\bibnamefont{Pauthenet}},\ }%
  \bibfield{journal}{%
  \bibinfo {journal} {Solid State Commun.}\ }%
  \textbf{\bibinfo {volume} {51}},\ \bibinfo {pages} {849} (\bibinfo {year}
  {1984})%
  \bibAnnoteFile{NoStop}{Twardowski2005}%
\bibitem{Wojcik2010}%
  \BibitemOpen
  \bibfield{author}{%
  \bibinfo {author} {\bibfnamefont{P.}~\bibnamefont{W{\'o}jcik}}, \bibinfo
  {author} {\bibfnamefont{B.~J.}\ \bibnamefont{Spisak}}, \bibinfo {author}
  {\bibfnamefont{M.}~\bibnamefont{Wo{\l}oszyn}},\ and\ \bibinfo {author}
  {\bibfnamefont{J.}~\bibnamefont{Adamowski}},\ }%
  \bibfield{journal}{%
  \bibinfo {journal} {Semicond. Sci. Technol.}\ }%
  \textbf{\bibinfo {volume} {25}},\ \bibinfo {pages} {125012} (\bibinfo {year}
  {2010})%
  \bibAnnoteFile{NoStop}{Wojcik2010}%
\bibitem{bookZiman2000}%
  \BibitemOpen
  \bibfield{author}{%
  \bibinfo {author} {\bibfnamefont{J.~M.}\ \bibnamefont{Ziman}},\ }%
  \emph{\bibinfo {title} {Electrons and phonons: the theory of transport
  phenomena in solids}}\ (\bibinfo {publisher} {Oxford University Press},\
  \bibinfo {year} {2000})%
  \bibAnnoteFile{NoStop}{bookZiman2000}%
\bibitem{Spisak2009}%
  \BibitemOpen
  \bibfield{author}{%
  \bibinfo {author} {\bibfnamefont{B.~J.}\ \bibnamefont{Spisak}}, \bibinfo
  {author} {\bibfnamefont{M.}~\bibnamefont{Wo{\l}oszyn}}, \bibinfo {author}
  {\bibfnamefont{P.}~\bibnamefont{W{\'o}jcik}},\ and\ \bibinfo {author}
  {\bibfnamefont{G.~J.}\ \bibnamefont{Morgan}},\ }%
  \bibfield{journal}{%
  \bibinfo {journal} {J. Phys.: Conf. Ser.}\ }%
  \textbf{\bibinfo {volume} {193}},\ \bibinfo {pages} {012130} (\bibinfo {year}
  {2009})%
  \bibAnnoteFile{NoStop}{Spisak2009}%
\bibitem{Weng2003}%
  \BibitemOpen
  \bibfield{author}{%
  \bibinfo {author} {\bibfnamefont{M.~Q.}\ \bibnamefont{Weng}}\ and\ \bibinfo
  {author} {\bibfnamefont{M.~W.}\ \bibnamefont{We}},\ }%
  \bibfield{journal}{%
  \bibinfo {journal} {J. Appl. Phys.}\ }%
  \textbf{\bibinfo {volume} {93}},\ \bibinfo {pages} {410} (\bibinfo {year}
  {2003})%
  \bibAnnoteFile{NoStop}{Weng2003}%
\bibitem{Spisak2005}%
  \BibitemOpen
  \bibfield{author}{%
  \bibinfo {author} {\bibfnamefont{B.~J.}\ \bibnamefont{Spisak}}, \bibinfo
  {author} {\bibfnamefont{A.}~\bibnamefont{Paja}},\ and\ \bibinfo {author}
  {\bibfnamefont{G.~J.}\ \bibnamefont{Morgan}},\ }%
  \bibfield{journal}{%
  \bibinfo {journal} {phys. stat. sol. (b)}\ }%
  \textbf{\bibinfo {volume} {242}},\ \bibinfo {pages} {1460} (\bibinfo {year}
  {2005})%
  \bibAnnoteFile{NoStop}{Spisak2005}%
\bibitem{Kim2007}%
  \BibitemOpen
  \bibfield{author}{%
  \bibinfo {author} {\bibfnamefont{K.~Y.}\ \bibnamefont{Kim}},\ }%
  \bibfield{journal}{%
  \bibinfo {journal} {J. Appl. Phys.}\ }%
  \textbf{\bibinfo {volume} {102}},\ \bibinfo {pages} {113705} (\bibinfo {year}
  {2007})%
  \bibAnnoteFile{NoStop}{Kim2007}%
\bibitem{Frensley1990}%
  \BibitemOpen
  \bibfield{author}{%
  \bibinfo {author} {\bibfnamefont{W.~R.}\ \bibnamefont{Frensley}},\ }%
  \bibfield{journal}{%
  \bibinfo {journal} {Rev. Mod. Phys.}\ }%
  \textbf{\bibinfo {volume} {62}},\ \bibinfo {pages} {745} (\bibinfo {year}
  {1990})%
  \bibAnnoteFile{NoStop}{Frensley1990}%
\bibitem{bookFerry2009}%
  \BibitemOpen
  \bibfield{author}{%
  \bibinfo {author} {\bibfnamefont{D.~K.}\ \bibnamefont{Ferry}}, \bibinfo
  {author} {\bibfnamefont{S.~M.}\ \bibnamefont{Goodnick}},\ and\ \bibinfo
  {author} {\bibfnamefont{J.~P.}\ \bibnamefont{Bird}},\ }%
  \emph{\bibinfo {title} {Transport in nanostructures}}\ (\bibinfo {publisher}
  {Cambridge University Press},\ \bibinfo {year} {2009})%
  \bibAnnoteFile{NoStop}{bookFerry2009}%
\bibitem{Chauvet2000}%
  \BibitemOpen
  \bibfield{author}{%
  \bibinfo {author} {\bibfnamefont{C.}~\bibnamefont{Chauvet}}, \bibinfo
  {author} {\bibfnamefont{E.}~\bibnamefont{Tourni{\'e}}},\ and\ \bibinfo
  {author} {\bibfnamefont{J.~P.}\ \bibnamefont{Faurie}},\ }%
  \bibfield{journal}{%
  \bibinfo {journal} {Phys. Rev. B}\ }%
  \textbf{\bibinfo {volume} {61}},\ \bibinfo {pages} {5332} (\bibinfo {year}
  {2000})%
  \bibAnnoteFile{NoStop}{Chauvet2000}%
\end{thebibliography}
\end{document}